\documentclass[a4paper,11pt]{article}
\pdfoutput=1 

\usepackage{jcappub} 

\usepackage[T1]{fontenc} 
\usepackage{footnote}

\usepackage{graphicx,ulem}
\usepackage{longtable}
\usepackage{float}
\usepackage{dcolumn}
\usepackage{graphics,epsfig}
\usepackage{amsmath,amssymb,latexsym,mathrsfs}
\usepackage{bm}
\usepackage{color}
\usepackage{color}
\usepackage{subfigure}

\usepackage{multirow}

\newcommand{\GeV}{\,\mathrm{GeV}}
\newcommand{\MeV}{\,\mathrm{MeV}}

\def\he4{$^4$He}
\newcommand{\Li}{$^7\mathrm{Li}$}
\newcommand{\Be}{$^7\mathrm{Be}$}
\def\h2{$^2$H}

\def\aap{\ref@jnl{A\&A}}                

\newcommand{\BR}{\mathrm{BR}}

\newcommand{\Gtog}{\Gamma_{\nu_S\to\nu\gamma}}

\begin{document}

\title{Breaking Be: a sterile neutrino solution to the cosmological lithium problem}

\date{\today}

\author{L.~Salvati$^1$, L. ~Pagano$^2$, M.~Lattanzi$^3$, M.~Gerbino$^{4,5}$, A.~Melchiorri$^1$}
\affiliation{$^1$ Physics Department and INFN, Universit\`a di Roma 
	``La Sapienza'', P.le\ Aldo Moro 2, 00185, Rome, Italy}
\affiliation{$^2$ Institut d'Astrophysique Spatiale, CNRS, Univ. Paris-Sud, Universit\'{e} Paris-Saclay, B\^{a}t. 121, 91405 Orsay cedex, France}	
\affiliation{$^3$ Dipartimento di Fisica e Scienze della Terra, Universit\`a di Ferrara and INFN, Sezione di Ferrara, Polo Scientifico e Tecnologico - Edificio C Via Saragat, 1, I-44122 Ferrara, Italy}
\affiliation{$^4$ The Oskar Klein Centre for Cosmoparticle Physics, Department of Physics, Stockholm University, AlbaNova, SE-106 91 Stockholm, Sweden}
\affiliation{$^5$ The Nordic Institute for Theoretical Physics (NORDITA), Roslagstullsbacken 23, SE-106 91 Stockholm, Sweden}
	
\emailAdd{laura.salvati@roma1.infn.it}
\emailAdd{lpagano@ias.u-psdu.fr}
\emailAdd{lattanzi@fe.infn.it}
\emailAdd{martina.gerbino@fysik.su.se}
\emailAdd{alessandro.melchiorri@roma1.infn.it}

\abstract{
The possibility that the so-called ``lithium problem'', i.e., the disagreement between the theoretical abundance predicted for primordial \Li\ assuming standard nucleosynthesis and the value inferred from astrophysical measurements, can be solved through a non-thermal Big Bang Nucleosynthesis (BBN) mechanism has been investigated by several authors. In particular, it has been shown that the decay of a MeV-mass particle, like, e.g., a sterile neutrino, decaying  after BBN not only solves the lithium problem, but also satisfies cosmological and laboratory bounds, making such a scenario worth to be investigated in further detail. In this paper, we constrain the parameters of the model with the combination of current data, including Planck 2015 measurements of temperature and polarization anisotropies of the Cosmic Microwave Background (CMB), FIRAS limits on CMB spectral distortions, astrophysical measurements of primordial abundances and laboratory constraints. We find that
a sterile neutrino with mass $M_S = 4.35 _{-0.17} ^{+0.13} \, \text{MeV}$ (at $95\%$ c.l.), a decay time $\tau _S = 1.8 _{-1.3} ^{+2.5} \cdot 10^5 \, \text{s}$ (at $95\%$ c.l.) and an initial density $\bar{n}_S/\bar{n}_{\text{cmb}}=1.7_{-0.6}^{+3.5} \cdot 10^{-4}$ (at $95\%$ c.l.) in units of the number density of CMB photons, perfectly accounts for the difference between predicted and observed $^7 \text{Li}$ primordial abundance. This model
also predicts an increase of the effective number of relativistic degrees of freedom at the time of CMB decoupling
$\Delta N_\text{eff}^\text{cmb}\equiv N_\text{eff}^\text{cmb} -3.046 = 0.34 _{-0.14} ^{+0.16}$ at $95\%$ c.l..
The required abundance of sterile neutrinos is incompatible with the standard thermal history of the Universe,
but could be realized in a low reheating temperature scenario.
We also provide forecasts for future experiments finding that the combination of measurements from the COrE+ and PIXIE missions will allow to significantly reduce the permitted region for the sterile lifetime and density.
}

\maketitle

\section{Introduction}\label{intro}

Big Bang Nucleosynthesis (hereafter BBN) successfully describes the production of the lightest nuclei in the first few minutes after the Big Bang,  \cite{Cyburt:2015mya}. 
A strong agreement has been found between the values of deuterium and helium abundances inferred from direct astrophysical observations (see e.g. Ref. \cite{Cooke:2013cba} for deuterium and Ref. \cite{Aver:2015iza} for helium) and the corresponding predictions based on the standard models of cosmology and particle physics. In fact, in this framework (that we shall refer to as ``standard BBN'') primordial abundances only depend on the baryon-to-photon ratio $\eta \equiv n_b/n_{\gamma}$ or, equivalently, on the baryon density $\omega_b \equiv \Omega_b h^2$. The latest Planck data release \cite{Adam:2015rua,Ade:2015xua} has shown the high degree of concordance between the 
theoretical abundances obtained from the baryon density inferred from Cosmic Microwave Background (hereafter CMB) anisotropies, and the abundances of $^4\text{He}$ and D measured through direct astrophysical observations. This concordance represents a great success for the Standard Model of Cosmology, making BBN one of the few probes of the primordial Universe.

On the other hand, a piece of the puzzle is still missing. In fact, there is a discrepancy between predicted and observed values of primordial $^7\text{Li}$, the so-called ``lithium problem" (see \cite{Fields:2011zzb,Cyburt:2015mya} for comprehensive reviews). In particular, the primordial abundance obtained from astrophysical observations is $\sim 3$ times smaller than theoretical predictions based on the Planck inferred value of the baryon density and D observations. 

The direct formation of mass-6 and mass-7 nuclei is suppressed during BBN, because of the absence of stable mass-5 nuclei. Thus $^7\text{Be}$ and $^7\text{Li}$ can be produced only through the fusion of lighter nuclei, in particular via $^3\text{He}(\alpha,\gamma)^7\text{Be}$ and $^3\text{H}(\alpha,\gamma)^7\text{Li}$.  At a later stage, $^7\text{Be}$ is converted into $^7\text{Li}$ via electron capture, in practice leaving lithium as the only mass-7 nucleus. Depending on the value of the baryon-to-photon ratio, the production of either  $^7\text{Li}$ or $^7\text{Be}$ is more efficient, and the final abundance of $^7\text{Li}$ will be mainly controlled by the direct fusion of lighter nuclei in the former case, or by  electron capture from $^7\text{Be}$ in the latter \cite{Steigman:2007xt}. 
In any case, given
the smallness of the reaction rates for the relevant processes with respect to the expansion rate of the Universe, only a small quantity of mass-7 nuclei ($\sim 10^{-10}$ with respect to hydrogen density) is produced.

Given that $^7\text{Li}$ is a weakly bound nuclide, different processes can affect its post-BBN evolution (see e.g. \cite{Cyburt:2015mya}), like stellar evolution (in particular during the asymptotic giant branch phase evolution of low-mass stars), or cosmic rays interactions with diffuse interstellar gas and neutrino spallation in supenovae "$\nu$ process". In general, $^7\text{Li}$ abundance is obtained from absorption spectra in metal-poor stars (Population II) atmosphere in the stellar halo of our Galaxy. 

Several evaluations of primordial $^7\text{Li}$ abundance from astrophysical observations are present in the literature that would lead to similar discrepancies with the theoretical value. In this paper, following the most recent BBN \cite{Cyburt:2015mya} and  \textit{Particle Data Group} reviews \cite{Fields:2014uja}, we decide to adopt the following value obtained from \cite{Sbordone:2010zi}, 

\begin{equation}
\left( \dfrac{^7\text{Li}}{\text{H}}  \right)_{\text{p}} = (1.6 \pm 0.3) \cdot 10^{-10} \, ,
\end{equation}

while, using the value $\Omega_bh^2 = 0.02222 \pm 0.00023$ measured by Planck \cite{Ade:2015xua}, the prediction of standard BBN is\footnote{In evaluating the total error for the predicted value we have considered both the statistical ($\sigma _{\text{stat}}=0.067$) and theoretical ($\sigma _{\text{th}}=0.32$) error. The latter is due to uncertainties in the interaction rates involved in $^7\text{Li}$ formation.} 
\begin{equation}
\left( \dfrac{^7\text{Li}}{\text{H}}  \right)_{\text{p,th}} = (4.51 \pm 0.33) \cdot 10^{-10} \, .
\end{equation}

Several proposals have been formulated to solve the lithium problem. 
As shown in \cite{Fields:2011zzb}, these possible solutions can be classified into different categories, depending on which part of the analysis is considered. It is possible to have astrophysical solutions that revise the measured primordial abundance. In fact, there could be systematics errors affecting current measurements or some errors related to the assumption that the plateau observed abundance  (the Spite plateau \cite{Spite:1982dd}) is the primordial one. However, it is important to consider that, even if these kind of astrophysical observations are in constant evolution, different results are stable in showing some amount of discrepancy between the observed and the theoretical value. It means that we can not rely only on an astrophysical solution for the lithium problem. Other types of solutions are related to nuclear physics, implying changes in the reaction rates for formation and destruction of mass-7 nuclei. It is important to consider that BBN calculation is very robust and that there have been strong improvements in the last years in constraining these rates, both with direct measurements and indirect analysis. For this reason, it does not seem to be a complete solution. 

If systematic errors in the astrophysical observations are negligible, we are left only with solutions beyond the Standard Model. One possibility is the decay of a heavy particles; in the last years several authors have explored this kind of solution to the cosmological lithium problem, such as in \cite{Jedamzik:2004er,Jedamzik:2005dh,Jedamzik:2007cp,Kusakabe:2007fu,Bailly:2008yy,Cyburt:2010vz,Kusakabe:2014ola,Ishida:2014wqa,Poulin:2015woa,Goudelis:2015wpa} (see also Ref. \cite{Fields:2011zzb} for a more extensive presentation of the previous literature on the subject, as well as for
discussion of solutions beyond the Standard Model that do not involve the decay of a heavy particle).
Following \cite{Poulin:2015woa}, in this paper we investigate the possibility to achieve consistency between the theoretical and observed values through  non-thermal primordial nucleosynthesis. In particular, we consider an injection of energy, due to the decay of a sterile neutrino, that will destroy part of $^7\text{Be}$ sometime after the BBN, preventing it to be converted into $^7\text{Li}$. We test this model against the most recent CMB data from Planck and astrophysical observations of primordial abundances of lithium and deuterium, adding also information on CMB spectral distortions and laboratory limits on active-sterile mixing. We also perform forecasts for future experiments.

We structure our work as follows: in sec. \ref{sec:model} we describe the assumed theoretical model and the current and future observables used to constrain it. In sec. \ref{sec:results} we present our results and, finally, in section \ref{sec:conclusions} we present our conclusions.

\section{Model}\label{sec:model}

As anticipated above, we consider non-thermal BBN as a possible solution to the lithium problem. Direct photo-disintegration of $^7 \text{Li}$ requires large amounts of energy injected into the plasma, which would impact the well-constrained abundances of lighter elements, as, for example, primordial deuterium that is in good agreement with the standard expectations. 
Instead, following Ref. \cite{Poulin:2015woa}, we consider the possibility that $^7 \textrm{Be}$ is partly destroyed,
before being converted to $^7\textrm{Li}$,  by photons associated to an electromagnetic cascade started by the decay of a sterile neutrino. The allowed energy range for photons involved in this process is quite narrow. 
The photon energy should be above the photo-disintegration threshold for $^7 \text{Be}$ of $1.6\,\mathrm{MeV}$. 
However, we want to avoid destroying deuterium, whose abundance is well constrained by observations. 
Then we expect to find that photon energies above the photo-disintegration threshold of deuterium,  $E_{\gamma} \le 2.2\,\mathrm{MeV}$, will be disfavoured. Since the whole process has to happen right after the epoch of standard BBN, and before beryllium is converted to lithium by electron capture, when the temperature of the plasma is below a few $\mathrm{keV}$, the decay of the sterile happens practically at rest.

In general, one can consider the following scenario: a generic species X with mass $M_X$ and lifetime $\tau_X$, decays injecting, on average, an energy $E_0$ in the plasma. The total energy released can be usefully expressed in terms of the parameter $\zeta_X=\bar{n}_X E_0/\bar{n}_\mathrm{cmb}$, where $\bar{n}_\mathrm{cmb}$ and $\bar{n}_X$
are the comoving densities of CMB photons and X's respectively, evaluated at a time $t \ll \tau_X$. Thus, $\zeta_X$ represents the average energy per photon that has been released after the decay of the whole X population. In the case of two-body decays of the form $X\rightarrow\gamma U$ ($U$ being a very light particle), given the considerations above, 
one can safely neglect thermal broadening and consider a monochromatic emission spectrum $p_\gamma=\delta(E_\gamma-E_0)$.

\subsection{Sterile neutrino}\label{sec:sterile}

In the following, we will focus on a specific implementation of this scenario, and assume that the $X$ particle is a sterile neutrino $\nu_S$. The existence of right-handed, sterile neutrino fields is not forbidden by any known symmetry of nature, and would allow for a Dirac mass terms for neutrinos, similarly to the SM charged fermions. Moreover, due to their gauge singlet nature,
they also allow for the presence of a bare Majorana mass term in the SM Lagrangian; the value of the corresponding mass scale is not bounded by any
theoretical consideration and has to be determined through observations. In fact, in see-saw models of neutrino mass generation  \cite{Minkowski:1977sc,Yanagida:1979as,Mohapatra:1979ia,GellMann:1980vs,Schechter:1980gr} the smallness of neutrino masses is obtained through a hierarchy between this scale and the vacuum expectation value of the Higgs field. 
 
We denote with $\theta_\alpha$ ($\alpha=e,\,\mu,\,\tau$) the mixing angles of the heavy, mostly sterile, state with the three active neutrinos.
In the following, we shall consider both cases of a Dirac and of a Majorana sterile neutrino. The two cases differ in the values of the decay rates for given mass and mixing angles (see below); moreover, some of the laboratory limits that we use in the following only apply if the sterile neutrino is a Majorana particle. The mass $M_S$ of the heavy neutrino, its comoving number density $\bar{n}_S$ at early (i.e., well before decay) times, the mixing angles $\theta_\alpha$, and its Dirac or Majorana nature completely specify the phenomenology of the model.

The mixing angles between active neutrinos and a heavy neutrino with mass $M_S > 1 \MeV$ have been constrained by many laboratory searches,
using different techniques that can be classified in three broad classes. The first includes searches for sterile neutrinos produced in accelerators. In principle, sterile neutrinos can be produced in high-energy collisions,
and would leave a signature if they decay inside the detector volume. The second technique is the study of decays in which neutrinos are produced in the final state; the presence of a heavy eigenstate would modify the kinematics of the decay. Mixing of the heavy state with the muon and electron neutrino
can be constrained by studying meson (pion and kaon) decays; in addition, the electronic mixing can be constrained by
tritium $\beta$ decay experiments. Finally, if neutrinos are Majorana particles, heavy states would contribute to the amplitude 
for neutrinoless double $\beta$ decay ($0\nu 2\beta$) proportionally to the $\theta_e$ mixing  angle, so that non-observation of 
$0\nu 2\beta$ decay allows to put upper limits on this parameter. Constraints on mixing angle from laboratory searches are nicely summarized
in Ref. \cite{Gelmini:2008fq}. In the following we shall use laboratory constraints as additional pieces of information to study the model under
consideration.

\begin{figure}[t!]
\centering
\includegraphics[scale=0.5]{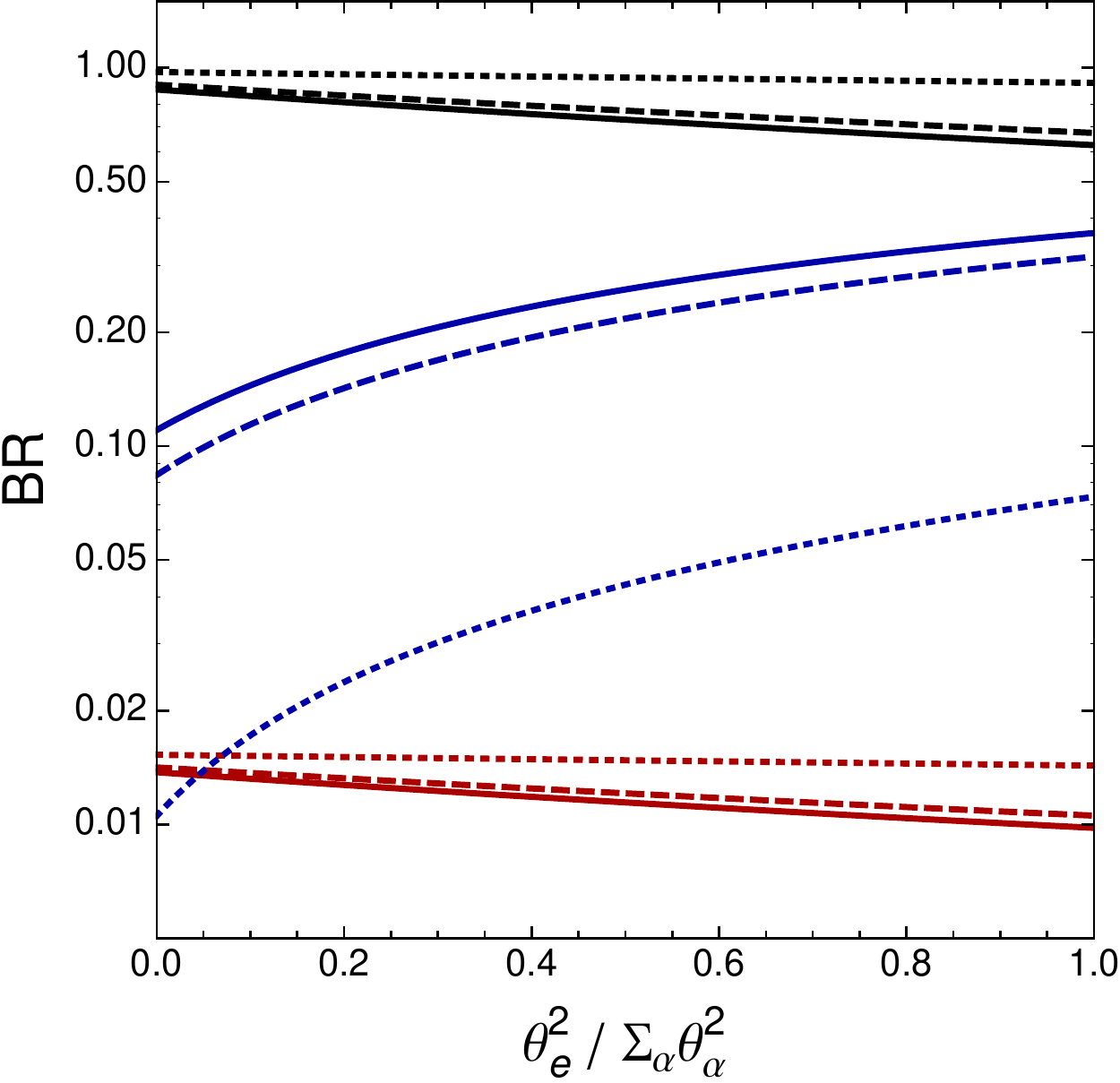}
\caption{\footnotesize{Branching ratios as function of $\theta_e^2 / \sum_\alpha \theta_\alpha^2$ for the three decay channels: $\nu _S  \rightarrow  3\nu$ in black, $\nu _S  \rightarrow  \nu e^+ e^- $ in blue and $\nu _S  \rightarrow  \nu \gamma $ in red. The solid, dashed and dotted lines are for $M_S \gg m_e$, $M_S = 8 m_e$ and $M_S = 3 m_e$, respectively.}}
\label{fig:BR}
\end{figure}

Sterile neutrinos mainly decay through neutral currents into three active neutrinos, $\nu _S  \rightarrow  3\nu$. 
Other relevant channels are those in which one light neutrino
in the final state is accompanied either by two charged leptons ($e^+ e^-$ for the range of masses under consideration), 
$\nu _S  \rightarrow  \nu \ell^+ \ell^- $, or by a photon, $\nu _S  \rightarrow  \nu \gamma $ .
The latter process, induced at the one-loop level,  is the one leading to photo-disintegration of \Be. The rates for these three processes are \cite{Denner:1992me,Johnson:1997cj,Lavoura:2003xp,Gorbunov:2007ak,Bezrukov:2009th}:
\begin{align}\label{eq:dec_ch}
& \Gamma _{\nu _S \,\, \rightarrow \,\, 3\nu}= \dfrac{1}{192 \pi^3} G_F^2 M_S^5 \sum_\alpha \theta^2 _{\alpha}  \notag \\
&\Gamma _{\nu_S \,\, \rightarrow \,\, \nu _{\alpha} e^+ e^-} = \dfrac{1}{192 \pi^3} G_F^2 M_S^5 \left[ \theta_e^2 f(x) + (\theta _{\mu}^2+\theta _{\tau}^2)g(x) \right]    \notag \\
&\Gamma _{\nu_S \,\, \rightarrow \,\, \nu \gamma} = \dfrac{9\alpha }{256 \pi^4} G_F^2 M_S^5 \sum_\alpha \theta^2 _{\alpha} \, ,
\end{align}
\noindent where $G_F$ is the Fermi coupling constant, $x \equiv m_e/M_S$, and the explicit expressions of $f(x)$ and $g(x)$  are given in Appendix \ref{app:decay_rates}. The above formulas are valid for Dirac neutrinos;
for Majorana neutrinos, they should be multiplied by a factor of 2, to account for charge-conjugated decay modes. 
The total decay rate $\Gamma_S $
 is dominated by the three-neutrino channel, but the electron/positron channel can also give a significant contribution (both are tree-level processes), while
 the one-loop radiative decay is subdominant. In the limit of massless electron, $M_s \gg m_e$ (i.e., $x \to 0$), the decay rate to $\nu e^+e^-$ is smaller that the rate to three neutrinos by a factor between $\sim 0.6$, for a sterile that only mixes with electron neutrinos, and $\sim 0.13$, for one that only mixes with $\mu$ and/or $\tau$ neutrinos. 
 This corresponds to branching ratios of $\sim 0.4$ and $\sim 0.1$ in the two cases, respectively. This numbers are further reduced for sterile masses in the MeV range due to kinematic suppression. The decay rate to photons, on the other hand, is $\sim 0.016$ times the decay rate to three neutrinos, independently of the mass and the mixing angles. We show the branching ratios in the various channels as functions of $\theta_e/\sum_\alpha \theta_\alpha^2$, for different values of $M_S$, in Figure \ref{fig:BR}.
 
It is useful to define average energy injection parameters, in analogy to $\zeta_X$ defined in the general case, one for each possible particle species in the final state. In other words, we would like to parameterize the average energy injected in the form of neutrinos (or photons, or $e^+ e^-$ pairs) after all the sterile neutrinos have decayed. 
 It is straightforward to show that this can be done by defining (we do not need to distinguish between different neutrino flavours, nor between electron and positrons)
\begin{equation}\label{eq:zeta_def}
\zeta_{S,i} \equiv \sum_j \BR_j \frac{\bar{n}_S \langle E \rangle_{i,j}}{\bar{n}_\mathrm{cmb}}, \quad ( i = \nu,\ \gamma,\ e) \, ,
\end{equation}
where the index $j$ runs over all processes with particle $i$ in the final state, $\BR_j \equiv \Gamma_j/\Gamma_s$ is the branching ratio of process $j$,
and $\langle E\rangle_{i,j}$ is the average energy of particle $i$ produced in process $j$.  

We also define for future convenience $\zeta_{S, \mathrm{EM}} \equiv
\zeta_{S, \gamma} + \zeta_{S,e}$.
Since the decay happens at rest, the total energy released in a given process is equal to $M_S$. For the leading decay mode, this clearly all goes to neutrinos, while for the radiative decay mode, the photon and the neutrino have an energy $E_0 = M_S/2$ each. The energy distribution in the $\nu_s \to \nu_\alpha e^+ e^-$ decay
depends on the ratio $m_e/M_S$; however, for $M_S \gg m_e$, the average energies of the final states are close to $M_S/3$ each. In fact, this is quite a good (at the 10\% level) approximation already at $M_S = 2.5\,\MeV$, as it can be seen looking, e.g., at Figure 16 of Ref. \cite{Ishida:2014wqa}, and we shall use that in the following.

Particles produced in sterile neutrino decays affect cosmological observables in several ways, that we shortly summarize here. Photons can disintegrate light nuclei and thus change light element abundances. Electromagnetic decay products induce spectral distortions in the CMB and increase the entropy of the cosmic plasma. Finally, light neutrinos increase the energy density of relativistic species,
changing the effective number of neutrino families $N_\mathrm{eff}$.
In the following subsections, we provide details about the impact of sterile neutrino decay on the main observables adopted in this work, namely the CMB and the light element abundances.

\subsubsection{Nonthermal Nucleosynthesis}\label{sec:NtBBN}
The first effect we want to describe is the one relevant for the proposed solution to the lithium problem. Through the $\nu _S  \rightarrow  \nu \gamma$ decay channel, photons with the right energy for the Be photo-disintegration are available. Provided that the decay happens after the BBN, enough Be can be depleted to lower the predicted abundance of lithium. 

The decay of the sterile contributes to the evolution of the photon distribution function $f_\gamma$ described by the usual Boltzmann equation. The spectrum of photons injected at redshift $z(t)$ by the decays is given by the source term $S_\gamma$:
\begin{equation}
S_\gamma(t)=n_S(t) \Gtog\, p_\gamma=\bar{n}_S(1+z(t))^3e^{-t/\tau_S} \Gtog p_\gamma \, ,
\end{equation}
where $\tau_S = \Gamma_S^{-1}$ is the lifetime of the sterile neutrino, and $p_\gamma$ is the spectrum of a single decay.
As pointed out in Ref. \cite{Poulin:2015woa}, if the decay happens at $T \lesssim$~few keV,
the energy of the primary injected photons is always below the pair-production threshold at the decay epoch,
so that there is no pair-production cutoff in the resulting electromagnetic cascade. This leads to a photon energy spectrum that is harder than the universal ``metastable'' spectrum typically used in the literature - in fact, one can argue that the processed spectrum of photons available for photodissociation is given by the decay spectrum, times a suppression due to the fact that some photons are ``lost'' in interactions with other components of the plasma. With this approximation, and assuming quasi-static equilibrium, $f_\gamma$ is simply given by the ratio between the source term and the rate $\Gamma_\gamma$ of the relevant interactions between photons and the other plasma components:  $f_\gamma = S_\gamma/\Gamma_\gamma$. The rate $\Gamma_\gamma$ takes into account all the interactions between photons and primordial plasma which contribute non-negligibly at the time $t$ and for the energy range considered in this process: Compton scattering over thermal electrons, scattering over CMB photons, pair production over nuclei and pair production over CMB photons. More details are reported in Appendix \ref{app:int_rates}.

Nuclear abundances during non-thermal BBN evolve according to the system of equations: 
\begin{eqnarray}\label{eq:balance}
\dfrac{dY_A}{dt} &=& \sum _T Y_T \int_0 ^{\infty} dE_{\gamma} f_{\gamma}(E_{\gamma}) \sigma_{\gamma + T \rightarrow A} (E_{\gamma})   \nonumber \\ 
 & -& Y_A \sum _P  \int_0 ^{\infty} dE_{\gamma} f_{\gamma}(E_{\gamma}) \sigma_{\gamma + A \rightarrow P} (E_{\gamma})
\end{eqnarray}
\noindent having defined $Y_A \equiv n_A/n_b$ as the abundance of nucleus $A$, and where $\sigma_{\gamma + T \rightarrow A}$ is the cross section
for production of $A$ through photodissociation over $T$, and $\sigma_{\gamma + A \rightarrow P}$ is the cross section for the analogous destruction channel. 

For the range of photon energies considered in our analysis, the relevant photodissociation processes, and the corresponding energy thresholds $E_{\mathrm{thr}}$, are \cite{Ishida:2014wqa}:
\begin{eqnarray}\label{eq:ntBBN}
^7\text{Be}(\gamma,^3\text{He})^4\text{He}  &\; \; \;\;\;\;\;& E_{\text{thr}}=1.5866 \,\text{MeV} \label{eq:ntBBN1} \\
 d(\gamma,n)p &\; \; \;\;\;\;\;& E_{\text{thr}}=2.2246 \,\text{MeV}    \label{eq:ntBBN2}\\
 ^7\text{Li}(\gamma,t)^4\text{He} &\; \; \;\;\;\;\;& E_{\text{thr}}=2.4670 \,\text{MeV} \, .  \label{eq:ntBBN3}
\end{eqnarray} 
\noindent We ignore the $^6 \text{Li}$ photo-disintegration (i.e. $^6 \text{Li}(\gamma,np)^4\text{He}$ with threshold $E_{\text{thr}}=3.6989 \,\text{MeV}$) since the initial abundance of this nucleus is negligible ($^6\text{Li}/\text{H} \sim 10^{-14}$). We also neglect the change in the abundances of the final products of these reactions, since the amount produced through photodissociation is much smaller than the yield of the same nucleus from the standard BBN phase, and only follow the evolution of \Li, $^7\mathrm{Be}$ and deuterium abundances. We only consider photodisintegration by primary photons produced in the decay; in principle, energetic photons could also be injected via the inverse Compton scattering of primary $e^+e^-$ off background photons. However a simple calculation shows
that the energy of these secondary photons is below the energy threshold for disintegration of $^7\text{Be}$ (the most weakly bound among light elements)
for the ranges of sterile mass and lifetime under consideration, 
and thus these processes can be safely neglected when calculating primordial abundances.

Integrating Eq. \eqref{eq:balance} over redshift, the initial-to-final abundance ratio for element $A$ can be written as: 
\begin{equation}\label{eq:ratio}
\text{ln} \left( \dfrac{Y_A(z_f)}{Y_A(z_i)} \right)= \int _{z_i} ^{z_f} \left.\dfrac{\bar{n}_\mathrm{cmb} \zeta _{S,\gamma} \sigma(E_0) e^{-1/(2H_r^0 \tau_S (1+z')^2)}}{E_0 H_r^0 \tau_S \Gamma_\gamma (E_0,z')}\right|_{E_0=M_S/2}dz' \, ,
\end{equation}
\noindent where we have assumed a monochromatic injection spectrum $p_\gamma = \delta(E-E_0)$, and used  
$H=H^0_r (1+z)^2$ (with $H_r^0 \equiv H_0 \Omega_\mathrm{rad}^{1/2}$) and $t =1/2H$ during the radiation-dominated era. 
The cross sections $\sigma$ for the reactions in Eq. \ref{eq:ntBBN1}-\ref{eq:ntBBN3} are taken from \cite{Cyburt:2002uv} for deuterium and from \cite{Ishida:2014wqa} for $^7\text{Be}$ and $^7\text{Li}$. 

\begin{figure}[t!]
\centering
\includegraphics[scale=0.5]{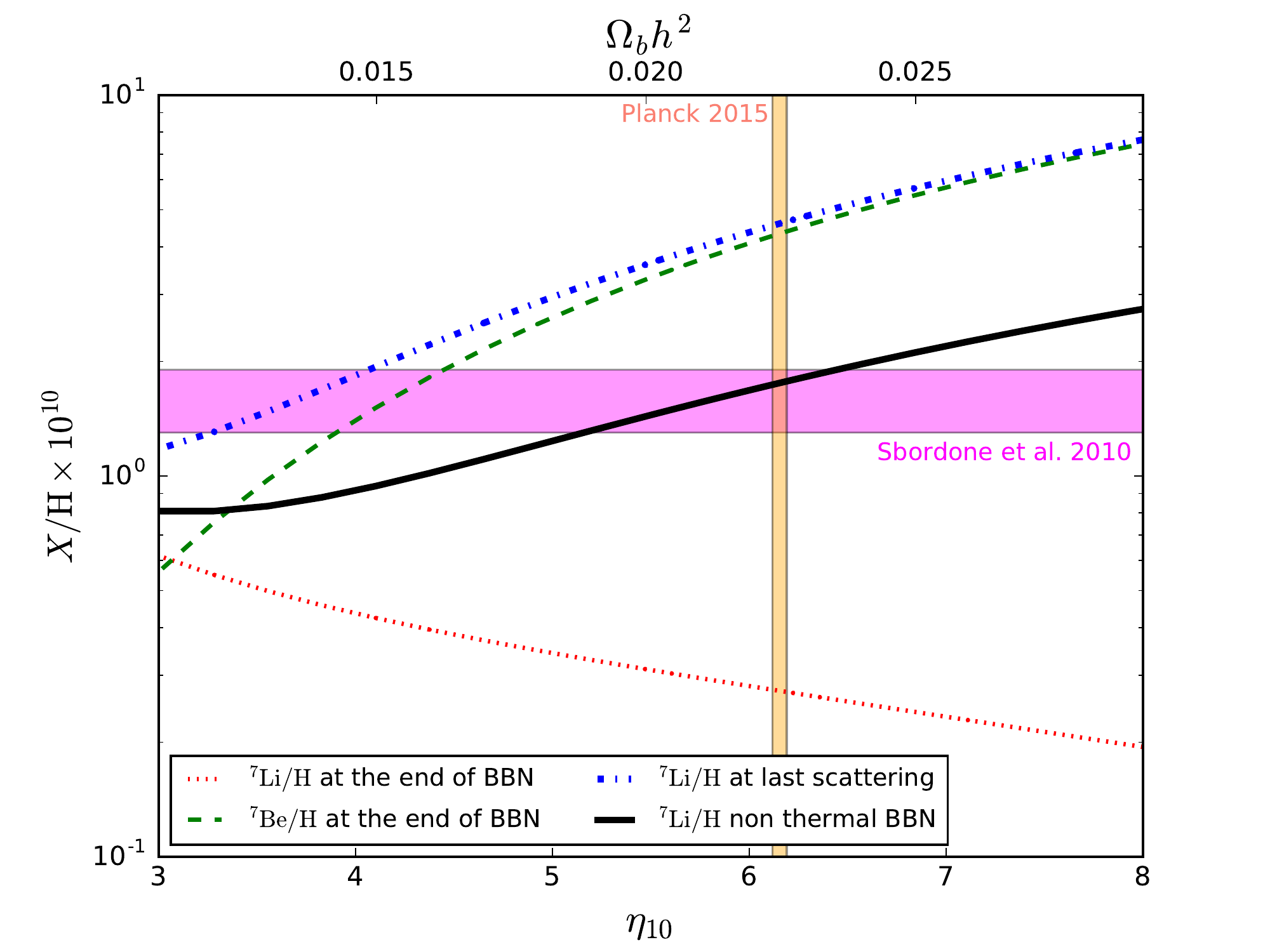}
\caption{\footnotesize{Abundances of mass-7 nuclei after BBN. The red dotted and the green dashed lines reproduce the abundances of \Li\ and \Be, respectively, right at the end of the thermal BBN phase. The blue dot-dashed line is the \Li\ abundance after all $^7\text{Be}$ is converted into $^7\text{Li}$ through electron capture, like in standard BBN. The black solid line is the final abundance of $^7\text{Li}$ assuming instead that $\sim65\%$ of $^7\text{Be}$ has been destroyed between the end of thermal BBN and the epoch of conversion through electron capture, as described in the text. The horizontal magenta band is the astrophysical measurement of $^7\text{Li}$, with the associated 1-$\sigma$ uncertainty \cite{Sbordone:2010zi}, and the vertical orange band is the estimate of $\Omega_bh^2$ from Planck 2015 \cite{Ade:2015xua}, with the associated 1-$\sigma$ uncertainty.}}
\label{fig:m7}
\end{figure}

In Figure \ref{fig:m7} we report theoretical predictions for mass-7 abundances, as functions of $\eta_{10}\equiv 10^{10} \eta$. The blue dot-dashed line is the total abundance of mass-7 nuclei at the end of thermal BBN. This is also equal to the final \Li\ abundance if all \Be\ nuclei are converted into \Li\ through electron capture, as in the standard scenario. In the model under consideration, however, a fraction of \Be\ is destroyed before electron capture becomes efficient. The black solid line is the final \Li\  abundance if we assume that $\sim 65\%$ of \Be\ has been photodisintegrated before being converted to \Li. 
It is seen that in this case the prediction for $^7\text{Li}$ abundance for the value of $\eta_{10}$ measured by Planck (vertical band) is now in agreement with astrophysical measurements, shown by the horizontal band.

\subsubsection{Spectral distortions}\label{sec:mu}
The energy release due to $\nu_S$ decays into photons and electrons/positrons can in principle produce spectral distortions in the CMB \cite{Chluba:2011hw}. At the redshift and for the typical lifetime of the $\nu_S$ we are considering, only $\mu$-type distortion can be created, since Compton scattering is still efficient in erasing $y$-type distortions. Following \cite{Poulin:2015woa}, the expected level of $\mu$ distortions for a sterile neutrino is
\begin{equation}\label{eq:mu}
\mu \simeq 8.01\cdot 10^2 \left( \dfrac{\tau _S}{1 \, \text{s}} \right)^{1/2}\left( \dfrac{\zeta _{S, \text{EM}}}{10^3\,\text{MeV}} \right) \mathcal{J} \, ,
\end{equation}
\noindent where the integral function $\mathcal{J}$ is taken from \cite{Chluba:2011hw}.

\subsubsection{Entropy variation}
Apart from spectral distortions, the energy release due to the electromagnetic decay of the sterile neutrino can be responsible for the increase of the photon entropy $S$. Since the decay takes place between the BBN and CMB epochs, if a considerable increase of entropy happens, this produces a change in the baryon-to-photon ratio $\eta$ so that $\eta_{\mathrm{BBN}}>\eta_{\mathrm{CMB}}$. 
However, there is a remarkable consistency between the values of $\eta$ inferred from BBN and CMB. We therefore expect the fractional difference in entropy due to energy injection from electromagnetic decay channels to be small, and we can approximate it as \cite{Feng_entropy}
\begin{equation}\label{eq:entropy}
\dfrac{\Delta S}{S} \simeq \text{ln} \left( \dfrac{S_\mathrm{after}}{S_\mathrm{before}} \right) =2.14 \cdot 10^{-4} \dfrac{\zeta _{S,\text{EM}}}{10^{-6} \text{MeV}} \left( \dfrac{\tau_S}{10^6 \, \text{s}} \right)^{1/2} \, ,
\end{equation}
\noindent where ``after'' and ``before'' refer to values after and before neutrino decay. Equation \eqref{eq:entropy} is valid in a radiation-dominated Universe.

\subsubsection{Contribution to neutrino background}\label{sec:Neff}
Since the main decay channel of the sterile neutrino is $\nu _S  \rightarrow  3\nu$, a relevant contribution of non-thermal neutrinos is produced. A non-negligible contribution also comes from the $\nu_S \,\, \rightarrow \,\, \nu _{\alpha} e^+ e^-$ channel. We can safely neglect the contribution from the radiative channel, considering its low branching ratio. This non-thermal term $N_{\mathrm{eff}}^{\mathrm{(nth)}} $  has to be included in the balance of the effective number of relativistic species $N_{\mathrm{eff}}$. Following \cite{Ishida:2014wqa}
\begin{eqnarray}\label{eq:neffcmb}
N_\mathrm{eff}(t) &=&\left(\frac{11}{4}\right)^{4/3}\left( \frac{T_\nu}{T}\right)^4  \Bigg[ N_\mathrm{eff}^{\mathrm{(std)}}  +  N_\mathrm{eff}^{\mathrm{(nth)}} \Bigg]   \notag \\
&=&\left(\frac{11}{4}\right)^{4/3}\left( \frac{T_\nu}{T}\right)^4  \Bigg[ N_\mathrm{eff}^{\mathrm{(std)}} + \frac{240\zeta(3)}{7\pi^4}\left(\frac{11}{4}\right)
\frac{\zeta_{S,\nu}}{\tau_s} \int_{t_\mathrm{in}}^t \frac{e^{-t'/\tau_s}}{T_\nu(t')}dt'\Bigg] \, .
\end{eqnarray}
\noindent Here, $N_{\mathrm{eff}}^{\mathrm{(std)}}=3.046$ is the standard number of relativistic species, $T$ and $T_\nu$ are the photon and thermal neutrino temperatures at the generic time $t$, respectively. This expression also takes into account the fact that, due to the entropy production discussed in the previous section,
the ratio between the temperatures of photons and relic neutrinos can be different from the standard value of $T_\nu/T = (4/11)^{1/3}$. 
In fact, before the decay of the sterile, one has $T_\nu/T = (4/11)^{1/3}$; however, immediately after the decay the photon temperature will 
be increased by a factor $(S_\mathrm{after}/S_\mathrm{before})^{1/3}$, while the temperature of neutrinos will stay unchanged, as no entropy is transferred to them. Thus in the end one has
\begin{equation}\label{eq:TTnu2}
T_\nu = \left(\frac{S_\mathrm{before}}{S_\mathrm{after}}\right)^{1/3}  \left(\frac{4}{11}\right)^{1/3} T \, .
\end{equation}

Finally, the integral in Eq. \eqref{eq:neffcmb}, can be approximated, for $t \gg \tau_S$ and for decays happening during the RD epoch (so that $T_\nu \propto t^{-1/2}$), as
\begin{equation}
\int_{t_\mathrm{in}}^t \frac{e^{-t'/\tau_S}}{T_\nu(t')}dt' \simeq \int_0^\infty \frac{e^{-t'/\tau_S}}{T_\nu(t')}dt' \simeq  \frac{\tau_S}{T_\nu(t_\mathrm{eff})} \, ,
\end{equation}
where $t_\mathrm{eff} \equiv (\pi/4) \tau_S$.

Putting everything together, we get
\begin{equation}
N_\mathrm{eff}^\mathrm{cmb} \equiv N_\mathrm{eff} (t \gg \tau_S) \simeq \left(\frac{S_\mathrm{before}}{S_\mathrm{after}}\right)^{4/3}  \Bigg[ N_\mathrm{eff}^{\mathrm{(std)}} + \frac{240\zeta(3)}{7\pi^2}\left(\frac{11}{4}\right) \frac{\zeta_{S,\nu}}{T_\nu(t_\mathrm{eff})} \Bigg] \, ,
\label{eq:neffcm2}
\end{equation}
where the factor in front is computed using Eq.~\eqref{eq:entropy} above, and the subscript ``CMB'' denotes that this is the value to which CMB observations are sensitive.
Computing this expression requires knowledge of the evolution of temperature (and thus of the cosmic scale factor) with time, that in turn requires to know $N_\mathrm{eff}$ itself, so that in fact $N_\mathrm{eff}$ implicitly appears on both sides of Eq.~\eqref{eq:neffcm2}. In order to find a solution to Eq.~\eqref{eq:neffcm2}, we make use of an iterative method that reaches full convergence within a few steps.

\section{Results}\label{sec:results}
We discussed in the previous section how the decay of a MeV neutrino affects (and as a consequence it can be constrained with) different observables. In this section after describing all the datasets that we have used to the purpose, we present the main results of our work.\\

\subsection{Description of the datasets}
As our baseline dataset, we consider the combination of CMB anisotropy data and direct astrophysical observations of primordial nuclei, in particular lithium and deuterium abundances. 
As CMB data, we use the temperature and polarization data from the latest Planck 2015 release \cite{Adam:2015rua,Aghanim:2015xee}, including
high-$\ell$ polarization. This is the dataset denoted as ``PlanckTTTEEE + lowP'' in the Planck collaboration papers; here we shall refer to it simply as ``Planck''.
The computation of the likelihood associated to the Planck data is performed using the publicly available Planck Likelihood Code 2.0 \cite{Aghanim:2015xee}.
Regarding lithium observations, we follow \cite{Cyburt:2015mya,Agashe:2014kda} and use, as the most reliable measure, the value reported in Ref. \cite{Sbordone:2010zi}, i.e. $^7\text{Li}/\text{H}=(1.6 \pm 0.3)\cdot 10^{-10}$. For deuterium abundance we adopt the most recent value from Ref. \cite{Cooke:2013cba}, i.e. $\text{D}/\text{H}=(2.53 \pm 0.04) \cdot 10^{-5}$ \footnote{During the review process of the present paper a new measure of the deuterium abundance was released by the same group \cite{Cooke:2016rky}, i.e. $\text{D}/\text{H}=(2.547 \pm 0.033) \cdot 10^{-5}$. Considering the theoretical error due to uncertainties on the nuclear rates, we do not expect any significant change in the results presented here.}. These measurements are denoted as ``Li'' and ``D'' in the following, and are implemented in the statistical analysis through Gaussian likelihoods on $^7\text{Li}/\text{H}$ and $\text{D}/\text{H}$. We also take into account the theoretical errors due to uncertainties on the nuclear rates involved in the BBN computation, in particular, for deuterium, we rely on the theoretical error stated in \cite{Ade:2015xua}, i.e. $\sigma (\text{D}/\text{H}) = 0.06 \cdot 10^{-5}$, obtained propagating the uncertainties quoted in \cite{Adelberger:2010qa}.

 The use of lithium observations is of course related to the main motivation of this work,
solving the lithium problem; on the other hand, the addition of deuterium astrophysical measurements is essential, since the same mechanism that might explain
the lithium discrepancy could also lead to photo-disintegration of deuterium, and spoil the excellent agreement between observations and the theoretical expectation.
In fact, as we shall see, including deuterium measurements will select values of the mass of the heavy neutrino below 4.4 MeV, so that the energy of the photon produced in the decay is below the threshold for deuterium photodissociation. Moreover, adding deuterium is also useful to obtain more stringent constraints on the number of relativistic degrees of freedom, since its abundance depends strongly on the total radiation content \cite{Cyburt:2015mya}.

The baseline data is complemented by considering additional pieces of information. As shown in sec. \ref{sec:mu}, the decay of a heavy neutrino can produce $\mu$-type distortions in the CMB energy spectrum. This leads us to consider the bound on $\mu$ distortions coming from the COBE-FIRAS measurements of the CMB frequency spectrum \cite{Fixsen:1996nj}, i.e. $\mu = (-1 \pm 4) \cdot 10^{-5} $, also implemented in the analysis through a Gaussian likelihood. Moreover, mixing angles of the sterile with active neutrinos are constrained by laboratory experiments, so it seems natural to consider also this information. In the range of masses we are interested in, the most stringent bounds come from the search for additional peaks in the spectrum of decays of mesons and $\tau$ leptons. For $\theta_e^2$ and  $\theta_{\mu}^2$ we have used results reported in Ref. \cite{Gelmini:2008fq} while for $\theta_{\tau}^2$ we have considered bounds from \cite{Helo:2011yg}. 
These bounds equally apply to Dirac and Majorana neutrinos, since they rely on purely cinematic arguments, and can be summarized, to a good approximation,
as $\theta_{e,\mu}^2 \lesssim 2.7 \times 10^{-5}$, and $\theta_\tau^2 \lesssim 8.8 \times 10^{-4}$. In the case of Majorana neutrinos, the electronic mixing angle 
is also bounded by searches for neutrinoless double beta decay \cite{Gelmini:2008fq}. This bound is much stronger than the one reported above, $\theta_e^2 \lesssim 10^{-7}$, and we shall use that when considering Majorana neutrinos. We implemented a likelihood function $\mathcal{L}(M_S,\theta_\alpha^2)$ based on probability distributions shown in 
 \cite{Gelmini:2008fq,Helo:2011yg} \footnote{The ``Lab'' likelihood is based on the figures 2 and 3 of \cite{Gelmini:2008fq} and on equation (43) of \cite{Helo:2011yg}.}.  The use of the likelihood from laboratory experiments is denoted with ``Lab'' in the following.
 
We also perform forecast for future CMB data, generating simulated datasets for the planned ground based SPT-3G telescope \cite{Benson:2014qhw}, for the future satellite missions COrE+\cite{CORE} and PIXIE \cite{Kogut:2011xw}.     

\subsection{Method}

In order to obtain parameter estimates, we use Monte Carlo Markov Chain (MCMC) package \texttt{cosmomc} \cite{Lewis:2002ah}, publicly available, which is built on a convergence diagnostic based on the Gelman and Rubin statistic. We use the February 2015 version that supports the Planck likelihood \cite{Aghanim:2015xee} and enforces efficiently the space sampling through a fast/slow parameters decorrelation \cite{Lewis:2013hha}. In our analysis we assume flatness and adiabatic primordial perturbations, and vary the six standard $\Lambda$CDM parameters: baryon density $\omega_b$, cold dark matter density $\omega_c$, sound horizon-to-angular diameter distance ratio at decoupling $\theta$, reionization optical depth $\tau$, scalar spectral index $n_s$, overall normalization of the spectrum $A_s$ at the pivot scale $k=0.05 \,\text{Mpc}^{-1}$.  We extend this minimal parameter space including the quantities relevant for our model: the sterile neutrino mass $M_S$, the $ \nu_S-\nu_{\alpha} $ mixing angles $\theta_{\alpha=e,\mu,\tau}$, the sterile neutrino initial comoving density $\bar{n}_S$. We treat the initial comoving density as a free parameter, in order not to assume any specific mechanism for the sterile neutrino production; however, in the Conclusions section, we will examine \textit{a posteriori} the implications of our findings with respect to the production mechanism. We assume a flat prior for the mass in the range $3.2 \div  11\,\MeV$, and logarithmic priors for the mixing angles and for the initial density.  We consider separately the two cases of Dirac or Majorana neutrinos. The contribution of the relic thermal neutrinos
to the number of relativistic degrees of freedom is fixed at the standard value of 3.046. We list in Table \ref{table:parameters} the above parameters, and some derived quantities useful to discuss the results of our analysis. 
We report our results in the form of 95\% bayesian credible intervals for the parameters.

We use \texttt{camb} \cite{Lewis:1999bs} to compute
spectra of CMB anisotropies given a particular realization of the model. As discussed in section \ref{sec:Neff}, 
we take into account that the radiation energy content associated to neutrinos is the sum of a standard term and a non-thermal term, generated by decay of the sterile neutrinos. The former is associated to the three active neutrinos and has the standard value of 3.046, while the latter is due to the sterile neutrino decay. We use Eqs. (\ref{eq:entropy}) and (\ref{eq:neffcm2}) to compute the total amount of relativistic degrees of freedom at each step of the MCMC. 

\begin{table}[H]
\scalebox{0.8}{
\centering
\begin{tabular}{|c|c|}
\hline
\multicolumn{2}{|c|}{base parameters} \\
\hline
\rule[-2mm]{0mm}{6mm}
$M_S$ & sterile neutrino mass \\
\hline
\rule[-2mm]{0mm}{6mm}
$\theta_e^2$ & $\nu_S-\nu_e$ mixing\\
\hline
\rule[-2mm]{0mm}{6mm}
$\theta_{\mu}^2$ & $\nu_S-\nu_{\mu} $ mixing \\
\hline
\rule[-2mm]{0mm}{6mm}
$\theta_{\tau}^2$ & $\nu_S-\nu_{\tau}$ mixing \\
\hline
\rule[-2mm]{0mm}{6mm}
$\bar{n}_S$ & initial comoving density of sterile neutrino \\
\hline
\end{tabular}}
\centering
\scalebox{0.8}{
\begin{tabular}{|c|c|c|c|}
\hline
\multicolumn{4}{|c|}{derived parameters} \\
\hline
\rule[-2mm]{0mm}{6mm}
$\tau_S$ & total decay time & $\zeta_{S,\gamma}$ & energy parameter for $\gamma$ \\
\hline
\rule[-2mm]{0mm}{6mm}
$\tau_{S,\gamma}$ & decay time for $\nu _S  \rightarrow  \nu \gamma$ & $\zeta_{S,\nu}$ & energy parameter for $\nu$\\
\hline
\rule[-2mm]{0mm}{6mm}
$\tau_{S,\nu}$ & decay time for $ \nu _S  \rightarrow  3\nu$ & $\zeta_{S,e}$ & energy parameter for $e^+ e^-$\\
\hline
\rule[-2mm]{0mm}{6mm}
$\tau_{S,e}$ & decay time for $\nu _S  \rightarrow  \nu _{\alpha} e^+ e^-$  & $\Theta ^2$ & total mixing angle \\
\hline
\end{tabular}}
\caption{\footnotesize{List of base parameters and derived parameters relevant for this model.}}
\label{table:parameters}
\end{table}

Similarly, we use the \texttt{PArthENoPE} code \cite{Pisanti:2007hk} to evaluate primordial yields of light elements, including those of  deuterium, $^7\text{Li}$ and $^7\text{Be}$, at the end of BBN ($T \sim \text{keV}$). These, assuming standard BBN, only depend on the baryon-to-photon ratio and on the number of relativistic degrees of freedom at the time of BBN; the latter is kept fixed to 3, while the former is computed from $\omega_b$, taking into account the entropy production associated to the sterile decay, see Eq. (\ref{eq:entropy}) (this is also indirectly relevant for the CMB, that is sensitive to the abundance of $^4$He).
Given the abundances at the end of the standard BBN phase, we use Equation (\ref{eq:balance}) to compute the abundances after the decay of the sterile,
to be compared with the values inferred from astrophysical observations.

Finally, we use Eq. (\ref{eq:mu}) to evaluate the expected amount of spectral distortions, and compare this with observations.

We conclude the description of the method by discussing the phenomenological consequences of the Dirac or Majorana nature of the sterile neutrino.
As discussed above, the decay rates for Majorana neutrinos are twice as large as those for Dirac neutrinos. If we do not consider limits from
laboratory experiments, our sensitivity to the mixing angles is only through the decay rates (or equivalently through the $\zeta_{S,i}$'s), as these enter directly in the calculation of the abundances, of the additional entropy and relativistic energy density, and of the spectral distortions.
Thus one can go from the Dirac to the Majorana case simply through the rescaling $\theta^2_\alpha \to \theta^2_\alpha/2$ (so that the $\Gamma$'s are equal
in the two cases), and the two cases will differ by a simple shift of the constraints in the parameter space. On the other hand, considering limits
from laboratory experiments breaks this degeneracy and could lead to non-trivial differences in the constraints between the two cases, especially
for what concerns the mixings; we shall see that this is indeed the case. Finally, we note that when laboratory constraints are not considered,
the data is not sensitive to $\theta_\mu$ and $\theta_\tau$ separately, but only on the combination $\theta_\mu^2 + \theta_\tau^2$, and 
the dimensionality of the parameter space can be reduced by one.

\subsection{Constraints from current data}
In this section we report results from current data.
Starting from the baseline combination (i.e. CMB plus measurements of primordial abundances), we add progressively the following datasets: FIRAS bounds on $\mu$-type spectral distortions \cite{Fixsen:1996nj}, and laboratory constraints on the mixing angles \cite{Gelmini:2008fq}. In doing so, we are able to gradually reduce the region of the parameter space allowed by data. The $95\%$ c.l. results for the relevant parameters are reported in Table \ref{table:Maj_Mvar} for the case of Majorana sterile neutrino and in Table \ref{table:Dir_Mvar} for the Dirac ones.

\begin{figure}[H]
\centering
\includegraphics[scale=0.40]{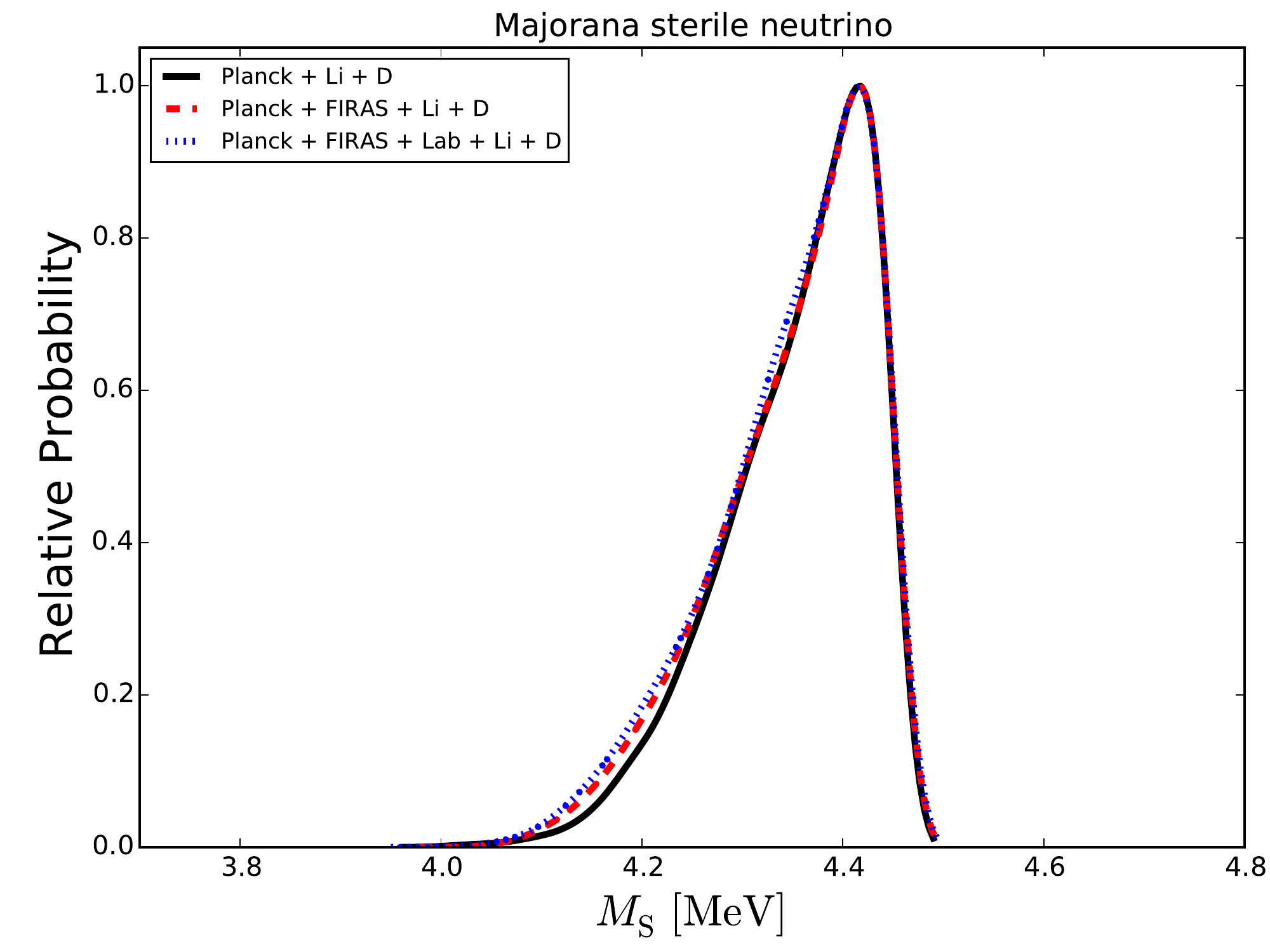}
\caption{\footnotesize{Posterior distribution for the mass of a heavy Majorana neutrino, for different dataset combinations.}}
\label{fig:Mvar}
\end{figure}

In Figure \ref{fig:Mvar} we show the posterior distribution for the sterile neutrino mass $M_S$, in the case of a Majorana neutrino; the corresponding result 
for a Dirac neutrino is practically identical, so we do not show it.
The gradual extension of dataset combinations does not impact the mass probability distributions. Indeed, the shape of the posteriors is completely dominated by primordial $^7\text{Li}$ abundance (which selects the lowest mass able to photodisintegrate \Be) and primordial deuterium abundance (which instead sets the highest allowed mass). In fact, the posterior distributions sharply peak around $4.4 \, \text{MeV}$, \textit{i.e.} the threshold for deuterium photo-disintegration.

\begin{center}
\begin{figure}[H]
\subfigure[{}]%
{\includegraphics[scale=0.38]{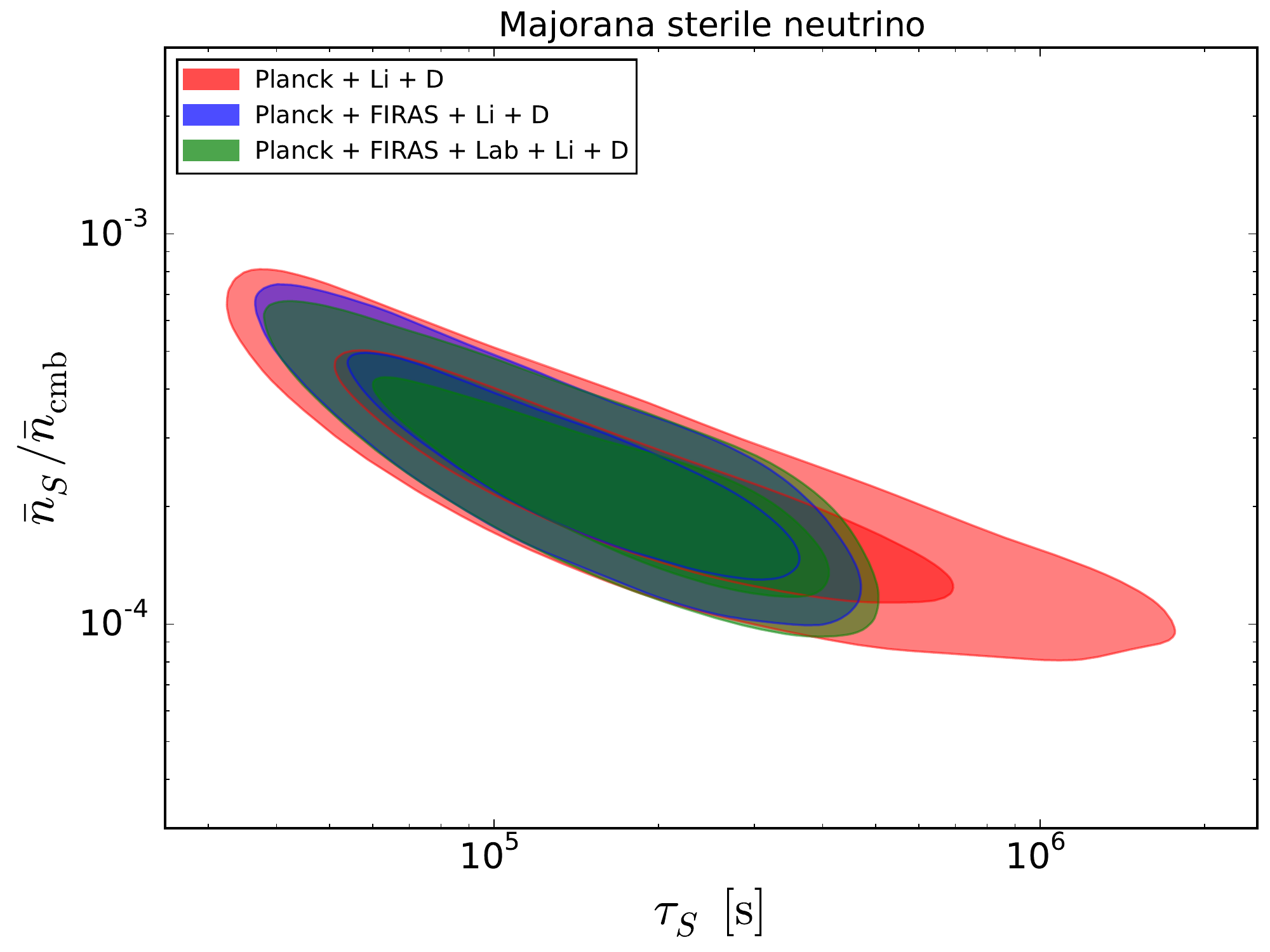}}
\subfigure[{}]%
{\includegraphics[scale=0.38]{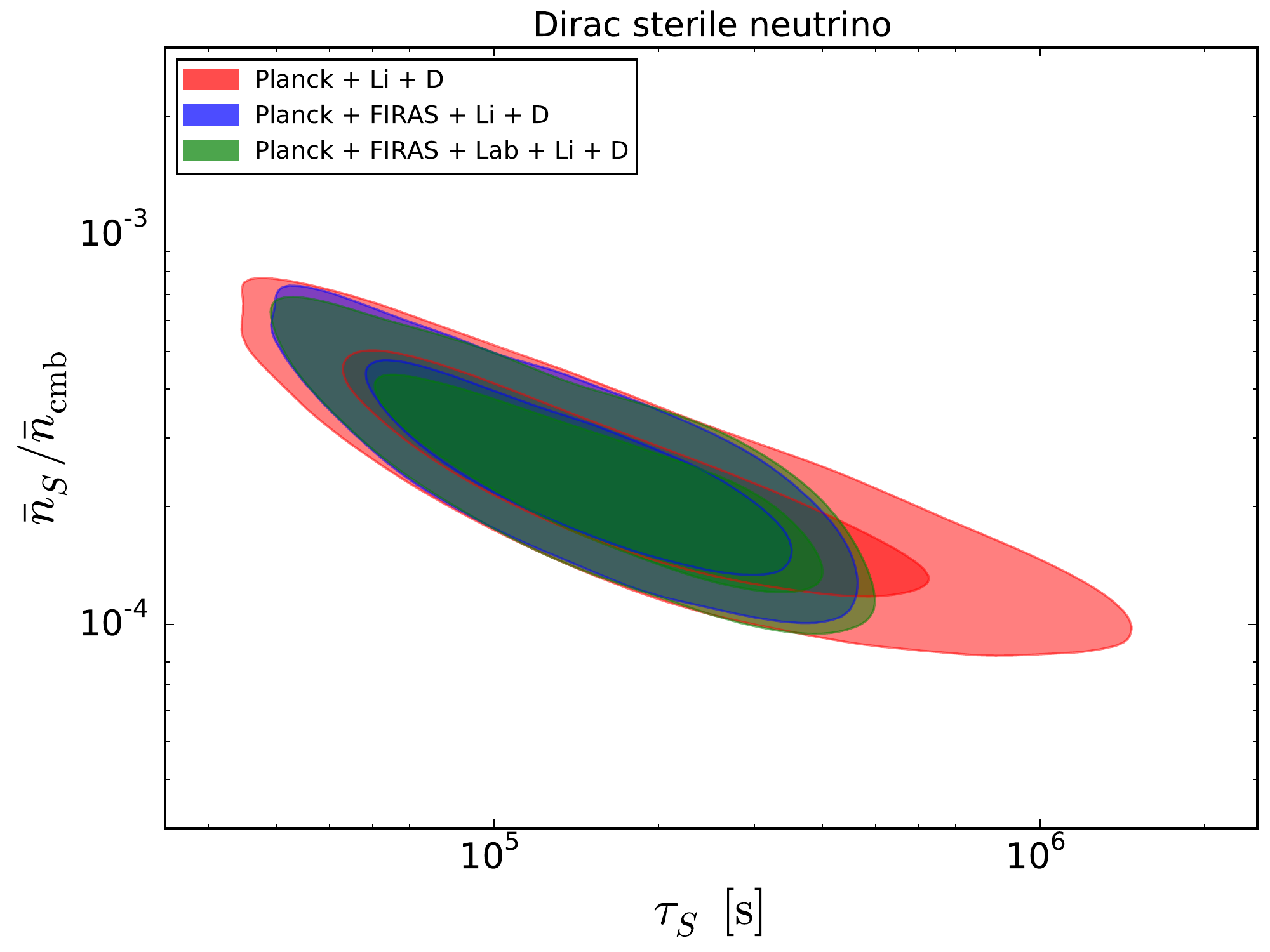}}
\caption{\footnotesize{Two-dimensional 68\% and 95\% credible regions for the decay time $\tau _S$ and comoving density $\bar{n}_S/\bar{n}_{\text{cmb}}$  of a heavy sterile neutrino, for the different combinations of datasets described in the text, in the case of a Majorana (a) and Dirac (b) neutrino.}}
\label{fig:tau-dens}
\end{figure}
\end{center}

In Figure \ref{fig:tau-dens}, we report two-dimensional probability distributions for the decay time $\tau _S$ and the density ratio $\bar{n}_S/\bar{n}_{\mathrm{cmb}}$, for different combinations of datasets. There are two different effects at play in shaping all the contours shown in the figure. The upper bound is determined by the CMB constraint on the total number of relativistic degrees of freedom $N_{\text{eff}}^{\text{cmb}}$: for a fixed value of $\tau _S$, the higher the comoving density, the greater the extra-radiation contribution coming from the sterile neutrino decay (see Eq. \ref{eq:neffcm2}, where the density enters through the $\zeta_{S,\nu}$). The lower bound can be set either by the astrophysical measurement of \Li\ or by a lower limit on $N_{\text{eff}}^{\text{cmb}}$. For what concerns the latter, it should be noted that $\bar{n}_S = 0$ (lying at negative infinity in the plot) corresponds to $N_{\text{eff}}^{\text{cmb}}=3.046$. Thus, in order to set a lower bound on $\bar{n}_S$ through a measurement of $N_{\text{eff}}^{\text{cmb}}$, the data
should provide evidence for $\Delta N_{\text{eff}}^{\text{cmb}} \equiv N_{\text{eff}}^{\text{cmb}}-3.046$ larger than zero. This is not the case from current data, so these do not help in this regard. On the other hand, a lower density results in a  less efficient \Be\ photo-disintegration process (see Eq. \ref{eq:ratio}) and in turn in a \Li\ abundance at variance with observations (i.e., exactly the lithium problem). Thus, in this case, too low values of $\bar{n}_S/\bar{n}_{\mathrm{cmb}}$ are excluded by astrophysical measurements of \Li\ abundance. 

The 2D constraints show an anticorrelation between the decay time and the density, related to the fact that both the final $^7\text{Li}$ abundance and the relativistic energy density can be kept approximately constant by decreasing the density while simultaneously increasing the decay time. 

Tighter constraints can be obtained by considering additional datasets. By adding the FIRAS bounds on $\mu$-type spectral distortions \cite{Fixsen:1996nj}, we can exclude higher values of the decay time $\tau _S$. This effect can be easily explained. A late decay time would produce later-time electromagnetic injection, with the plasma being unable to efficiently thermalize, the net result being the enhancement of $\mu$-type spectral distortions, as it can be seen from Eq. \ref{eq:mu}.
On the other hand, inclusion of laboratory bounds on the mixing angles \cite{Gelmini:2008fq} does not substantially change the constraints with respect to the Planck + Li + D + FIRAS case.

The limits in the ($\bar{n}_S/\bar{n}_{\mathrm{cmb}}$, $\tau _S$) plane translate to $\Delta N_{\text{eff}}^{\text{cmb}}= 0.34 _{-0.14} ^{+0.16}$ at a 95\%  c.l. in the Majorana case, for Planck + FIRAS + Lab + Li + D.

As expected, the constraints in the $(\tau_S,\,\bar{n}_S/\bar{n}_{\mathrm{cmb}})$ plane do not change between the Dirac and Majorana case, when only information from cosmological observables is used. However, as noted above, direct constraints on the mixing angles from laboratory experiments do not affect the results, that thus continue to be the same for the Dirac and Majorana cases. This was not obvious \textit{a priori} and is telling us that the information we have on the mixing angles (different in the two cases) is not constraining enough to further exclude values of the decay time $\tau_S$ that would otherwise be allowed by cosmological observations, especially those on spectral distortions.

\begin{center}
\begin{figure}[H]
\subfigure[{}]%
{\includegraphics[scale=0.38]{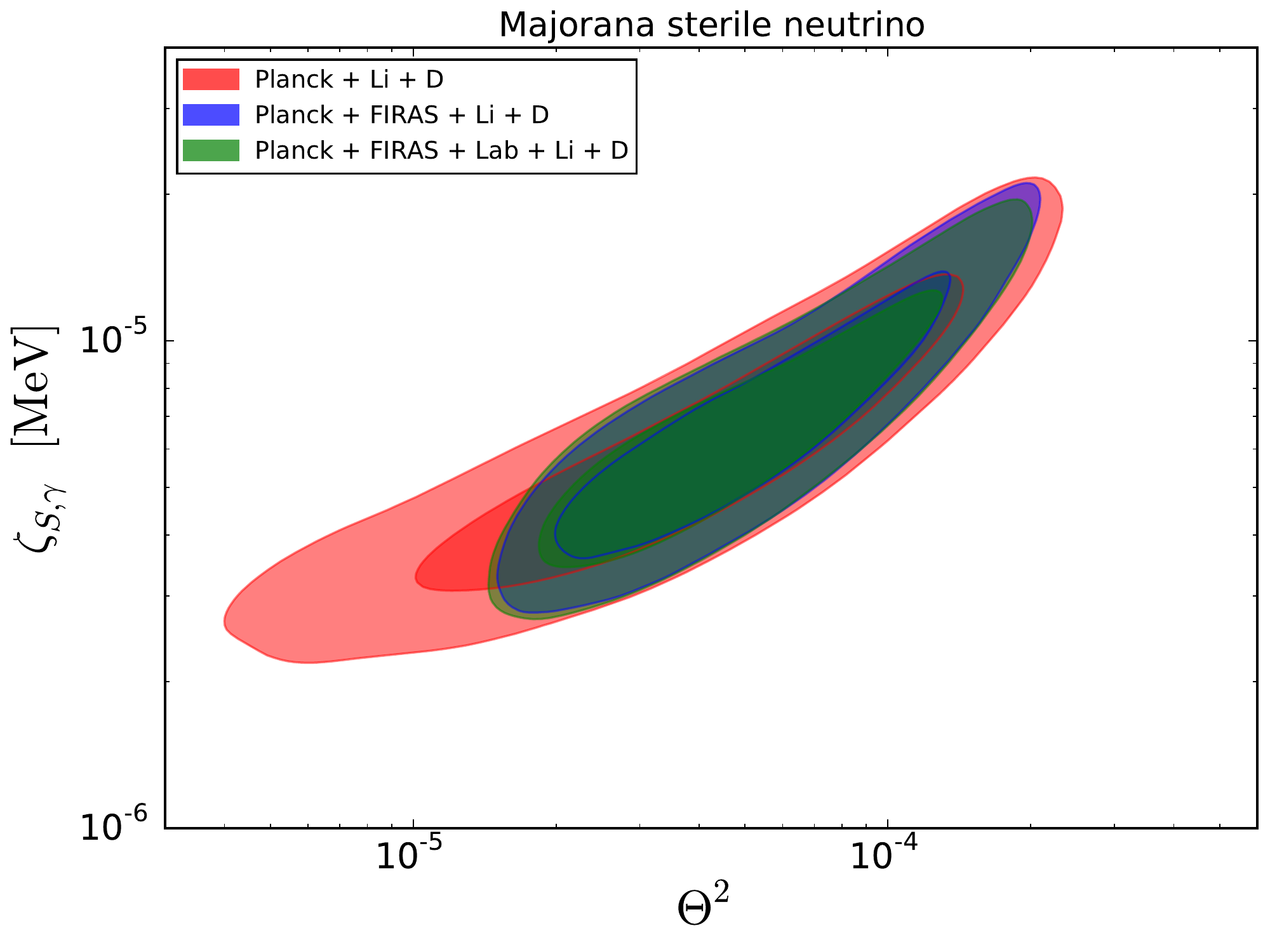}}
\subfigure[{}]%
{\includegraphics[scale=0.38]{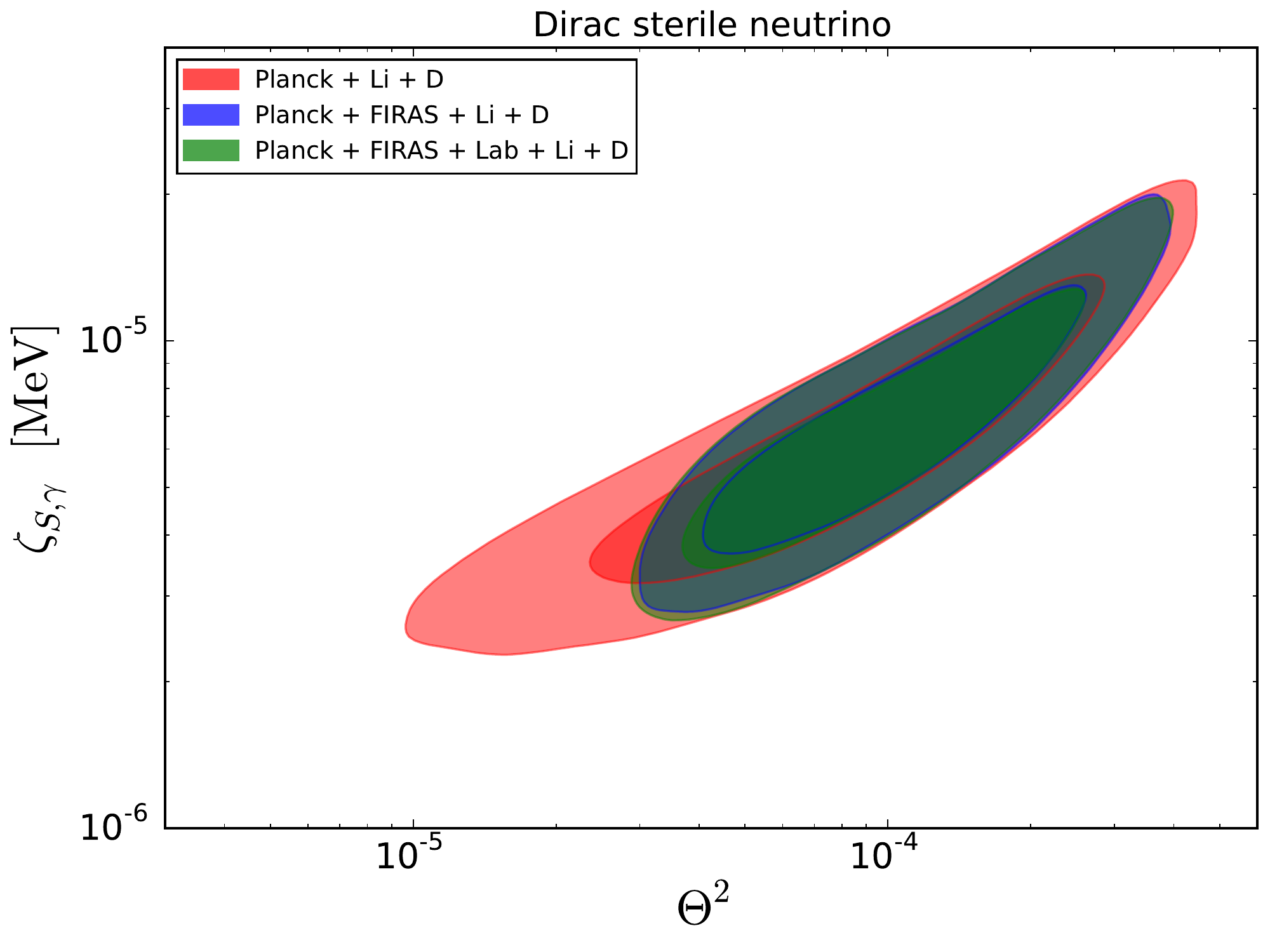}}
\caption{\footnotesize{Two-dimensional 68\% and 95\% credible regions for the total active-sterile mixing angle $\Theta ^2$ and energy parameter for photons $\zeta _{S,\gamma}$, for the different combinations of datasets described in the text, in the case of a Majorana (a) and Dirac neutrino (b).}}
\label{fig:theta-zetag}
\end{figure}
\end{center}

On the other hand, the difference between Dirac and Majorana is relevant when the results are expressed in terms of the mixing angles. In Figure \ref{fig:theta-zetag}, the two-dimensional distributions for the total mixing angle $\Theta ^2$ and the photon energy parameter $\zeta_{S,\gamma}$ are depicted. When looking at the ($\Theta ^2$, $\zeta_{S,\gamma}$) plane, the addition of FIRAS constraints on $\mu$-type spectral distortions cuts the distribution for lower values of $\Theta ^2$, as expected from Eq \ref{eq:dec_ch}, given that $\Theta ^2 \propto 1/\tau_S$. For the same reason, the correlation between the two parameters is now positive. For these parameters, conversely to what shown in Figure \ref{fig:tau-dens}, there is a difference between Majorana and Dirac case. In particular, the contours appear shifted, reflecting the different proportionality between interaction rates and $\theta_{\alpha}$, as mentioned in Sec. \ref{sec:sterile}.

\begin{center}
\begin{figure}[t!] 
\subfigure[{}]%
{\includegraphics[scale=0.38]{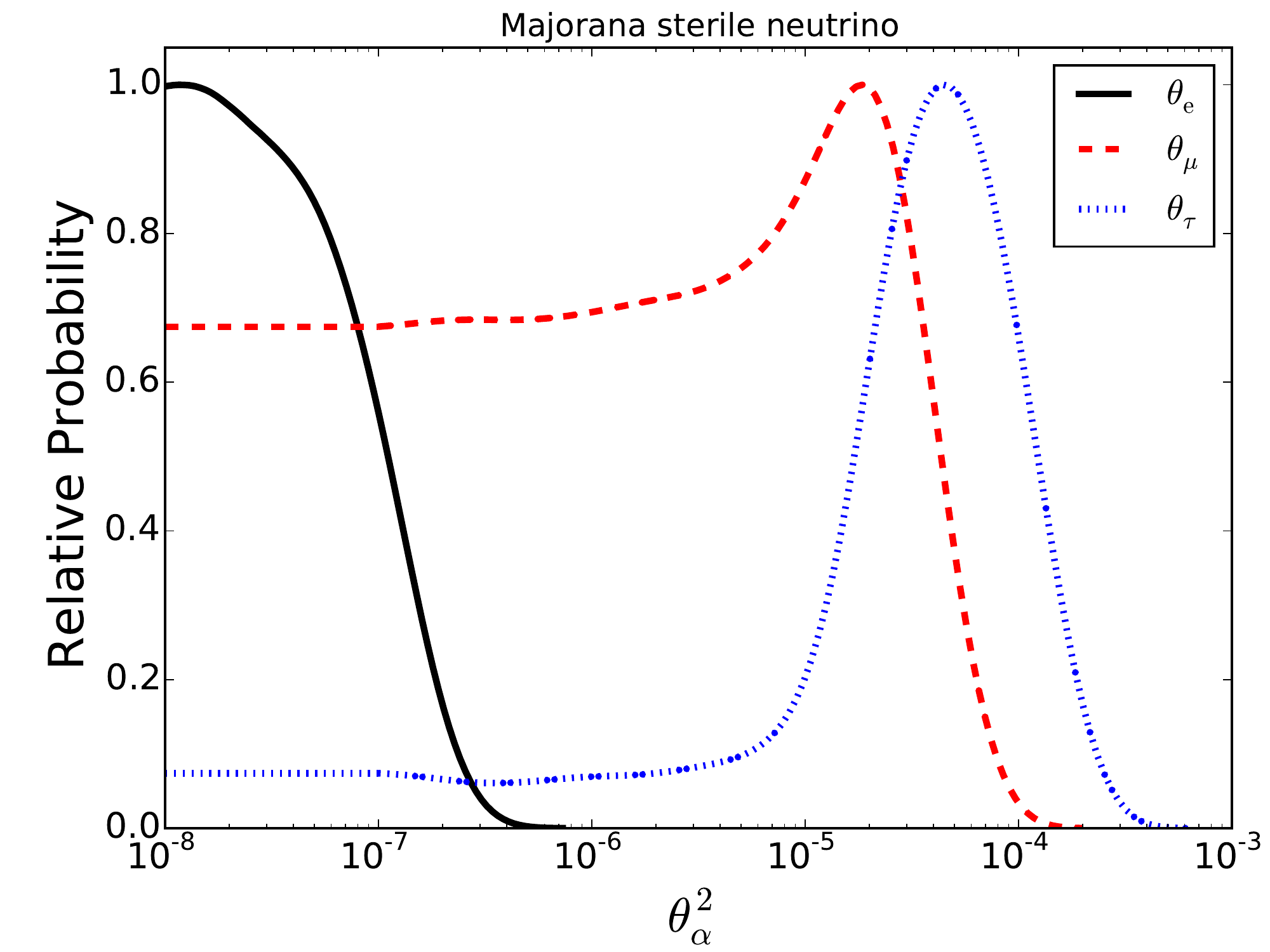}}
\subfigure[{}]%
{\includegraphics[scale=0.38]{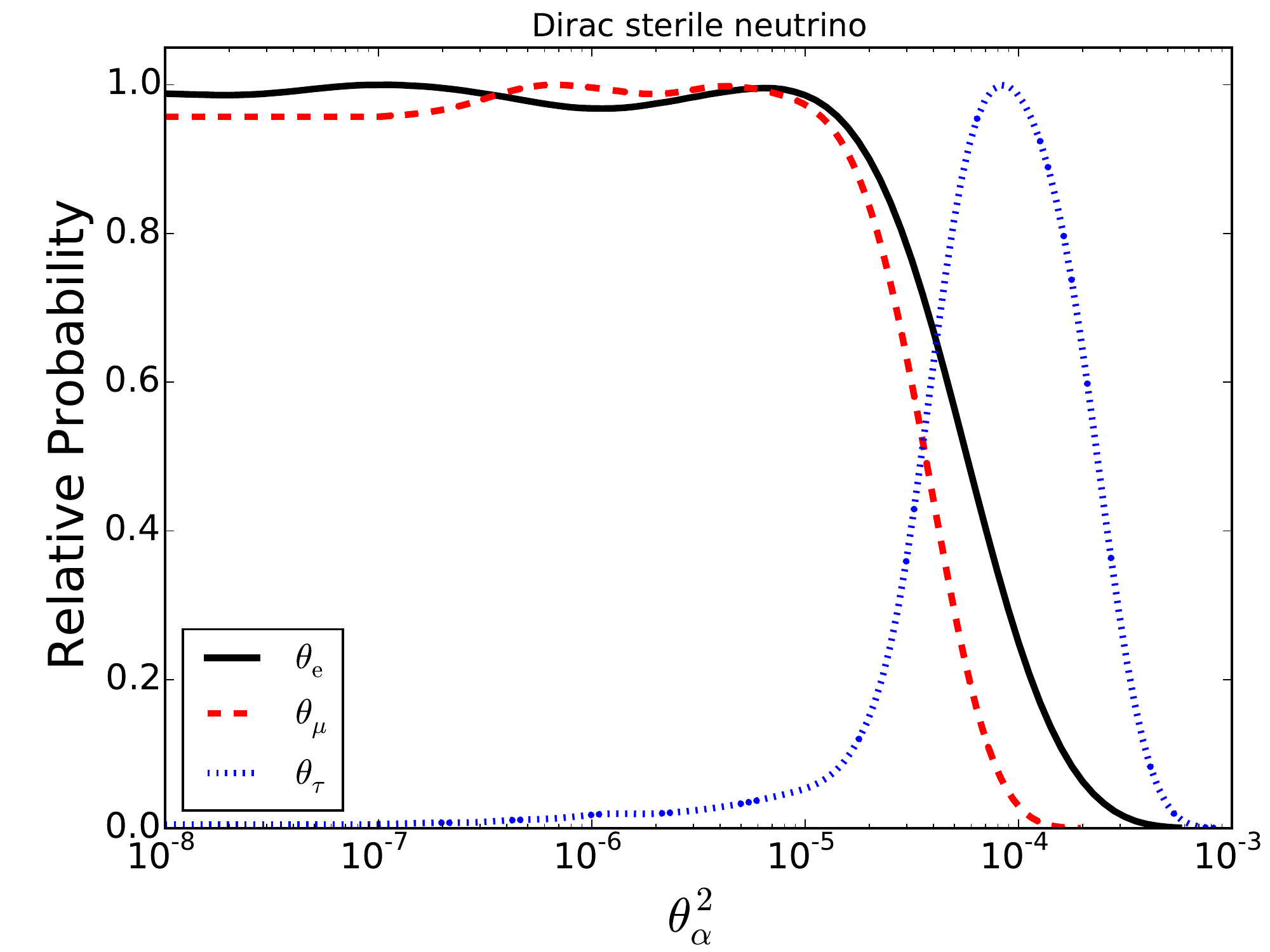}}
\caption{\footnotesize{Posterior distributions for single flavour mixing angles ($\theta_e$, $\theta _{\mu}$, $\theta _{\tau}$) for the datasets combination Planck + FIRAS + Lab + Li + D (described in the text), in case of Majorana sterile neutrino (a) and Dirac sterile neutrino (b)}}\label{fig:theta-Mvar}
\end{figure}
\end{center}

The posterior distributions for the single mixing angles are shown in Figure \ref{fig:theta-Mvar}. We note that, since some of these posterior distributions do not 
vanish at the prior boundary, and this boundary is not set by a physical requirement, the 95\% credible intervals for the mixing angles reported in
Tables \ref{table:Maj_Mvar} and \ref{table:Dir_Mvar} somehow depend on the choice
of where to cut the distribution at small values of the angles. This situation is often encountered in practice, when dealing
with logarithmic priors on parameters that the data cannot exclude being equal to zero. In this case, direct inspection
of one-dimensional posterior distributions like those shown in Figure \ref{fig:theta-Mvar} gives a more robust assessment of the information
provided by the data. We see that in the Majorana case, the sterile is mostly mixed with $\tau$ and $\mu$ neutrinos, since the mixing with electron neutrinos 
is bound to be very small by the non-observation of neutrinoless $\beta$ decay. The mixing with $\nu_\tau$ is further preferred to the mixing with $\nu_\mu$ because
mixing angles $\sim \mathrm{few}\times 10^{-5}$ are required to solve the lithium problem, and these are only marginally allowed by kinematic bounds on $\mu$ mixing, but are fully allowed by the looser bounds on $\tau$ mixing. In the Dirac case, the bound on $\theta_e$ and $\theta_\mu$ are similar, both coming from kinematic measurements, 
and the sterile is mostly mixed with $\tau$ neutrinos, for the same reason as in the Majorana case, while the mixing with the other active neutrino flavours is lower by nearly an order of magnitude. 

\begin{figure}[t!]
\centering
\includegraphics[scale=0.50]{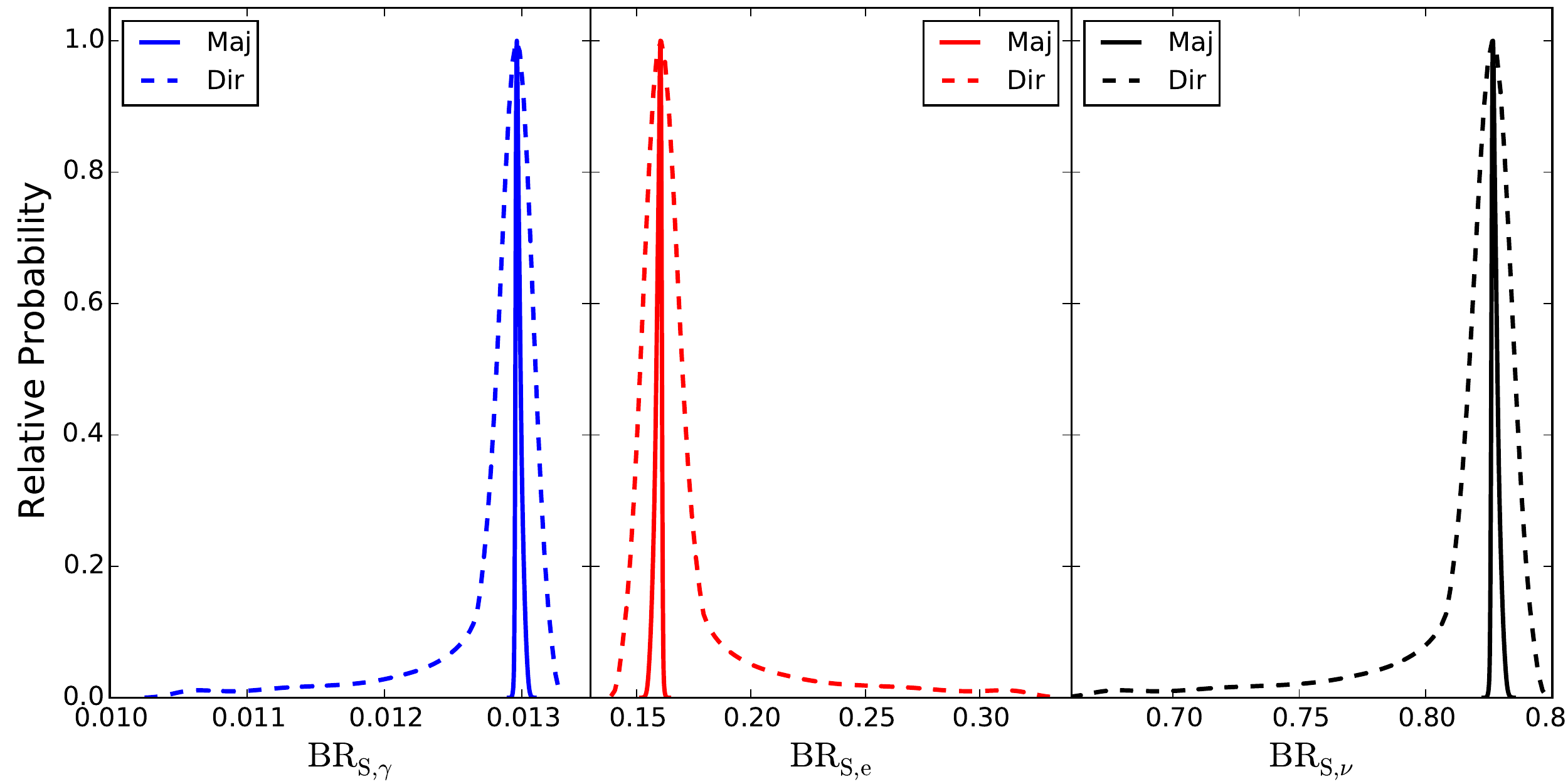}
\caption{\footnotesize{Branching ratio posterior distributions for different decay channels for Majorana (solid line) and Dirac (dashed line) sterile neutrinos. The color code is consistent with the one used in Figure \ref{fig:BR}.}}
\label{fig:BR_tot}
\end{figure}

In Figure \ref{fig:BR_tot} we show the branching ratio posterior distributions for the different decay channels, for both Dirac and Majorana sterile neutrinos. The posterior widths reflect the theoretical predictions on the branching ratios shown in Figure \ref{fig:BR} and results on the mixing angles shown in Figures \ref{fig:theta-Mvar}. In fact, since laboratory experiments constraint  $\theta _e ^2 / \sum _{\alpha} \theta ^2 _{\alpha} \simeq 0$ for a Majorana neutrino, the branching ratios are only functions of the sterile neutrino mass and the allowed parameter space is narrower than the Dirac case (see also Appendix \ref{app:decay_rates} and in particular Equation \ref{eq:decay2leptons}).

The results reported so far answer the question of what are the limits on the parameters of the model, assumed to be true.
Another important question is whether the model represents an improvement with respect to standard $\Lambda$CDM and, in the case of a positive answer,
whether this improvement, likely due to a better agreement of the model predictions with observations of  \Li\ abundance, comes at the expense
of the agreement with CMB measurements. A precise answer would require to perform  model comparison in a Bayesian framework;
for the purpose of this paper, we will content ourselves by doing a simple $\chi^2$ test analysis.
We compare the mean $\chi^2$ associated to each dataset, also taking into account the additional parameters (and consequently the degrees of freedom) with respect to the standard $\Lambda$CDM model. We observe no significant difference for the CMB data, finding $\chi^2_{\text{CMB}}=12949$ for the $\Lambda$CDM + $\Delta N_\text{eff}$ (with $\Delta N_\text{eff}>0$) model fitted to Planck data only, $\chi^2_{\text{CMB}}=12953$ for the $\Lambda$CDM case (assuming SBBN and fitting both Planck and \Li\ abundance data) and $\chi^2_{\text{CMB}}=12951$ for the model employed in this paper (Majorana case, using Planck + FIRAS + Lab + Li + D).\\
Concerning $\chi^2_{\text{Li}}$, we obtain a dramatic improvement: for the $\Lambda$CDM case, we have $\chi^2_{\text{Li}}=85$, while for our model we obtain $\chi^2_{\text{Li}}=1.4$. The relevance of this finding is more evident if one considers that the model described here only accounts for 5 additional degrees of freedom with respect to the standard $\Lambda$CDM model.\\
Similar results also apply to the Dirac case. Thus we conclude that a cosmological model with a decaying heavy neutrino can simultaneously solve the lithium problem and provide an excellent fit to CMB data, performing in this last respect not worse than the $\Lambda$CDM model.

\begin{table}
\begin{center}
\scalebox{0.72}{
\begin{tabular}{|c|c|c|c|c|c|c|c|}
\multicolumn{8}{c}{\textbf{Majorana sterile neutrino}}\\
\hline
\rule[-2mm]{0mm}{6mm}
\multirow{2}*{\textbf{Dataset}}& $M_S$ & $\bar{n}_S/\bar{n}_{\text{cmb}}$ & $ \tau _{S,\text{tot}}$ &  $\zeta_{S,\gamma}$ &$\zeta_{S,\nu}$& $ \zeta _{S,e}$ &\multirow{2}* {$\Theta ^2 \cdot 10^4$}  \\
\rule[-2mm]{0mm}{6mm}
 & $[\text{MeV}]$ & $10^4$ & $10^{-5}[\text{s}]$ &  $10^5 [\text{MeV}]$ &$10^3 [\text{MeV}]$& $10^4 [\text{MeV}]$ &   \\
\hline
\rule[-2mm]{0mm}{6mm}
Planck + Li + D& $ [4.20,4.47]$  & $[0.9,5.5] $ &$[0.4,9.6] $ & $[0.3,1.5] $ &$ [0.4,2.0]$ & $[0.4,3.7] $& $[0.1,1.8] $ \\
\hline
\rule[-2mm]{0mm}{6mm}
Planck + FIRAS + Li + D&  $ [4.19,4.47]$  &$ [1.2,5.7]$ &$[0.5,4.0]$ & $[0.3,1.5] $& $[0.4,2.1] $&$[0.6,3.6] $  & $[0.2,1.6] $\\
\hline
\rule[-2mm]{0mm}{6mm}
Planck + FIRAS + Lab +& $[4.19,4.47] $ &$[1.1,5.2] $ & $[0.5,4.4] $& $[0.3,1.4] $&$[0.4,2.0] $ & $[0.5,2.4] $ & $[0.2,1.5] $\\
\rule[-2mm]{0mm}{6mm}
 +  Li + D&  &  &   & & & &$\theta ^2 _e < 1.1 \cdot10^{-7} $ \\
\rule[-2mm]{0mm}{6mm}
 & &   &   & & & &$\theta ^2 _{\mu} < 3.4 \cdot10^{-5} $\\
\rule[-2mm]{0mm}{6mm}
 & &   &   & & & & $\theta ^2 _{\tau} =[2.2 \cdot 10^{-7},1.8 \cdot 10^{-4}]$ \\
 \hline
\end{tabular}}
\caption{\footnotesize{95\% credible intervals for the sterile neutrino parameters, for the different combinations of datasets.}}
\label{table:Maj_Mvar}
\end{center}
\end{table}

\begin{table}
\begin{center}
\scalebox{0.72}{
\begin{tabular}{|c|c|c|c|c|c|c|c|}
\multicolumn{8}{c}{\textbf{Dirac sterile neutrino}}\\
\hline
\rule[-2mm]{0mm}{6mm}
\multirow{2}*{\textbf{Dataset}}& $M_S$ & $\bar{n}_S/\bar{n}_{\text{cmb}}$ & $ \tau _{S,\text{tot}}$ &  $\zeta_{S,\gamma}$ &$\zeta_{S,\nu}$& $ \zeta _{S,e}$ &\multirow{2}* {$\Theta ^2 \cdot 10^4$}  \\
\rule[-2mm]{0mm}{6mm}
 & $[\text{MeV}]$ & $10^4$ & $10^{-5}[\text{s}]$ &  $10^5 [\text{MeV}]$ &$10^3 [\text{MeV}]$& $10^4 [\text{MeV}]$ &   \\
\hline
\rule[-2mm]{0mm}{6mm}
Planck + Li + D& $ [4.19,4.47]$  & $[1.0,5.4] $ &$[0.4,8.7] $ & $[0.3,1.4] $ &$ [0.4,2.0]$ & $[0.5,3.7] $& $[0.2,3.4] $ \\
\hline
\rule[-2mm]{0mm}{6mm}
Planck + FIRAS + Li + D&  $ [4.18,4.47]$  &$ [1.2,5.5]$ &$[0.5,3.9]$ & $[0.3,1.5] $& $[0.4,2.0] $&$[0.6,3.7] $  & $[0.4,3.0] $\\
\hline
\rule[-2mm]{0mm}{6mm}
Planck + FIRAS + Lab +& $[4.17,4.47] $ &$[1.1,5.2] $ & $[0.5,4.2] $& $[0.3,1.4] $&$[0.4,2.0] $ & $[0.5,2.5] $ & $[0.3,2.9] $\\
\rule[-2mm]{0mm}{6mm}
+ Li + D&  &  &   & & & & $\theta ^2 _e < 3.3 \cdot 10^{-5} $ \\
\rule[-2mm]{0mm}{6mm}
 & &   &   & & & & $\theta ^2 _{\mu} < 2.9 \cdot 10^{-5}] $\\
\rule[-2mm]{0mm}{6mm}
 & &   &   & & & &$\theta ^2 _{\tau} =[0.11,4.46]\cdot 10^{-4}$ \\
 \hline
\end{tabular}}
\caption{\footnotesize{95\% credible intervals for sterile neutrino parameters, for the different combinations of datasets.}}
\label{table:Dir_Mvar}
\end{center}
\end{table}

\subsection{Forecasts on future data}

In this section we perform forecasts on future experiments, mainly focusing on CMB experiments. The aim of this analysis is to understand to what extent
future datasets will improve current constraints on the model.

Concerning measurements of CMB anisotropies, both in temperature and polarization, we consider forthcoming results from the ground based SPT-3G telescope \cite{Benson:2014qhw}, and a future CMB space mission such as COrE+ \cite{CORE}. The SPT data are always used in combination
with the present Planck data. For what concerns spectral distortions, we take into account the expected sensitivity of the future PIXIE mission \cite{Kogut:2011xw}. 
We perform the forecasts analysis only for the Majorana sterile neutrino case since we reasonably expect to obtain similar results for the Dirac case, motivated by what discussed in the previous section. For the forecast analysis we assume white noise, gaussian beam and negligible residuals due to foregrounds and systematics. As our fiducial model, we use the best-fit values from Planck + FIRAS + Lab + Li + D dataset combination, reported in Table \ref{table:fid}.

\begin{table}[t!]
\begin{center}
\scalebox{0.8}{
\begin{tabular}{|c|c||c|c||c|c||c|c||c|c||c|c|}
\multicolumn{12}{c}{base parameters} \\
\hline
\multicolumn{2}{|c|}{$\Omega_bh^2$} & \multicolumn{2}{c|}{$0.022619 $} & \multicolumn{2}{|c|}{$\Omega_c h^2$} & \multicolumn{2}{c|}{$0.1236 $} & \multicolumn{2}{|c|}{$100\, \theta_{MC}$} & \multicolumn{2}{c|}{$ 1.04063$}\\
\hline
 \multicolumn{2}{|c|}{$\tau$}& \multicolumn{2}{c|}{$ 0.0977$} & \multicolumn{2}{|c|}{${\rm{ln}}(10^{10} A_s)$} &\multicolumn{2}{c|}{$3.1358 $} & \multicolumn{2}{|c|}{$n_s$}&\multicolumn{2}{c|}{$ 0.9759$} \\
\hline
\multicolumn{4}{|c|}{$M_S$} &\multicolumn{2}{c|}{$4.40 $} & \multicolumn{4}{|c|}{$\bar{n}_S/\bar{n}_{\text{cmb}} $}& \multicolumn{2}{c|}{$0.00017$}  \\
\hline
\multicolumn{2}{|c|}{$\theta ^2 _e$} &\multicolumn{2}{c|}{$1.5 \cdot 10^{-8} $} & \multicolumn{2}{|c|}{$\theta ^2 _{\mu}$}&\multicolumn{2}{c|}{$1.3 \cdot 10^{-7} $} &\multicolumn{2}{|c|}{ $\theta ^2 _{\tau}$} &\multicolumn{2}{c|}{$4.0 \cdot 10^{-5} $ }\\
\hline
\multicolumn{12}{c}{derived parameters} \\
\hline
 \multicolumn{2}{|c|}{$\tau_S$} & \multicolumn{2}{c|}{$1.80 \cdot 10^5$}& \multicolumn{2}{|c|}{$N_{\text{eff}}^{\text{cmb}}$} & \multicolumn{2}{c|}{$ 3.321$} & \multicolumn{2}{|c|}{$\mu$} &\multicolumn{2}{c|}{$3.0 \cdot 10^{-7}$} \\
\hline
\end{tabular}}
\caption{\footnotesize{Fiducial parameter values used in the forecast analysis.}}
\label{table:fid}
\end{center}
\end{table}

We start by considering the combination of future CMB anisotropies data with current spectral distortions limits and laboratory measurements of mixing angles (Planck + SPT-3G + FIRAS + Lab and COrE+ + FIRAS + Lab). 
First of all, we assess the ability of future CMB data to constrain the model with minimal input from astrophysical observations of nuclear abundances. 
These are needed to constrain the sterile neutrino mass, so we fix it to be around the largest value allowed by present
observations, just below the threshold for deuterium photo-disintegration, \textit{i.e.} $M_S = 4.4 \, \text{MeV}$.

\begin{center}
\begin{figure}[t!]
\subfigure[{}]%
{\includegraphics[scale=0.38]{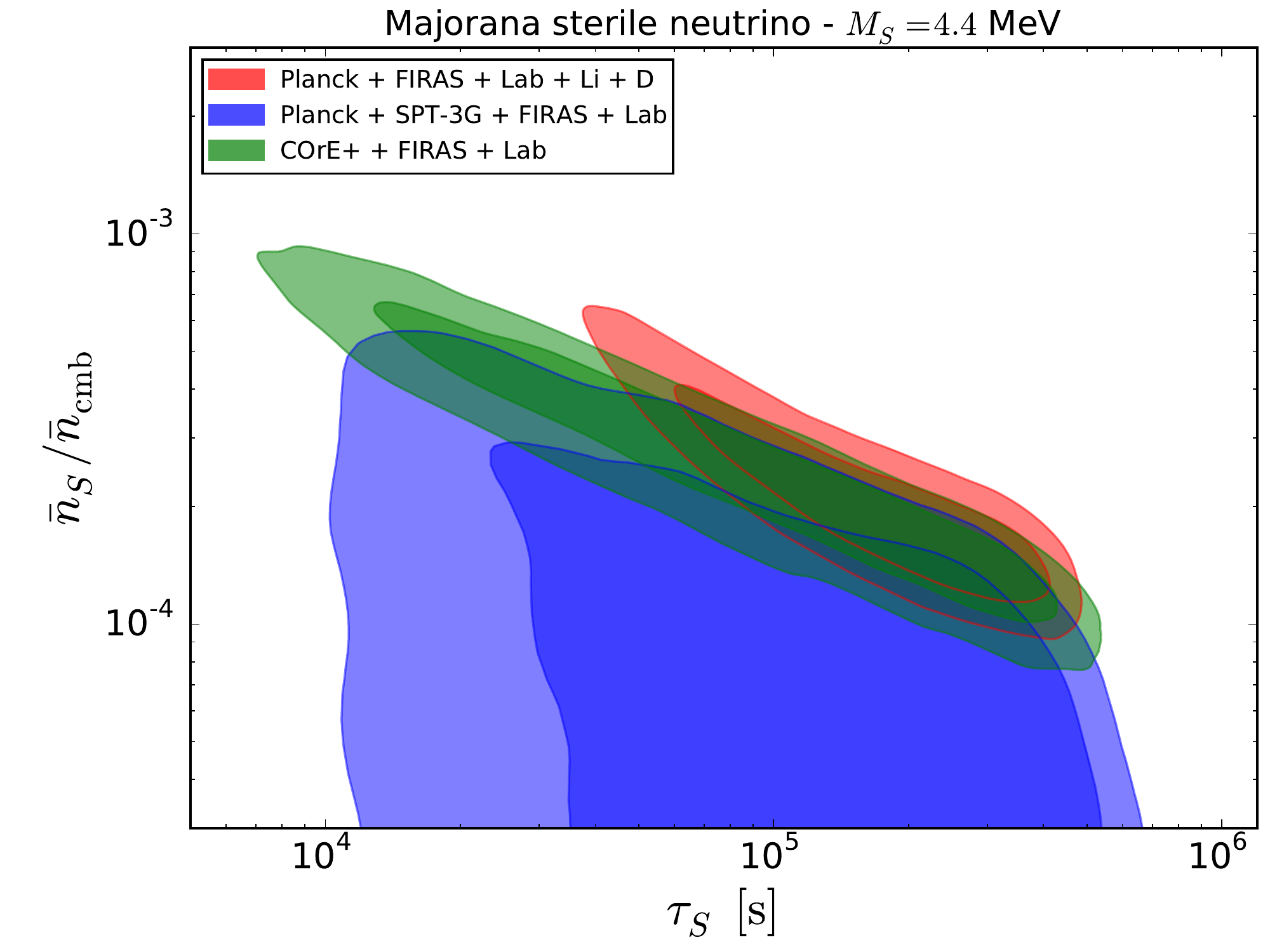}}
\subfigure[{}]%
{\includegraphics[scale=0.38]{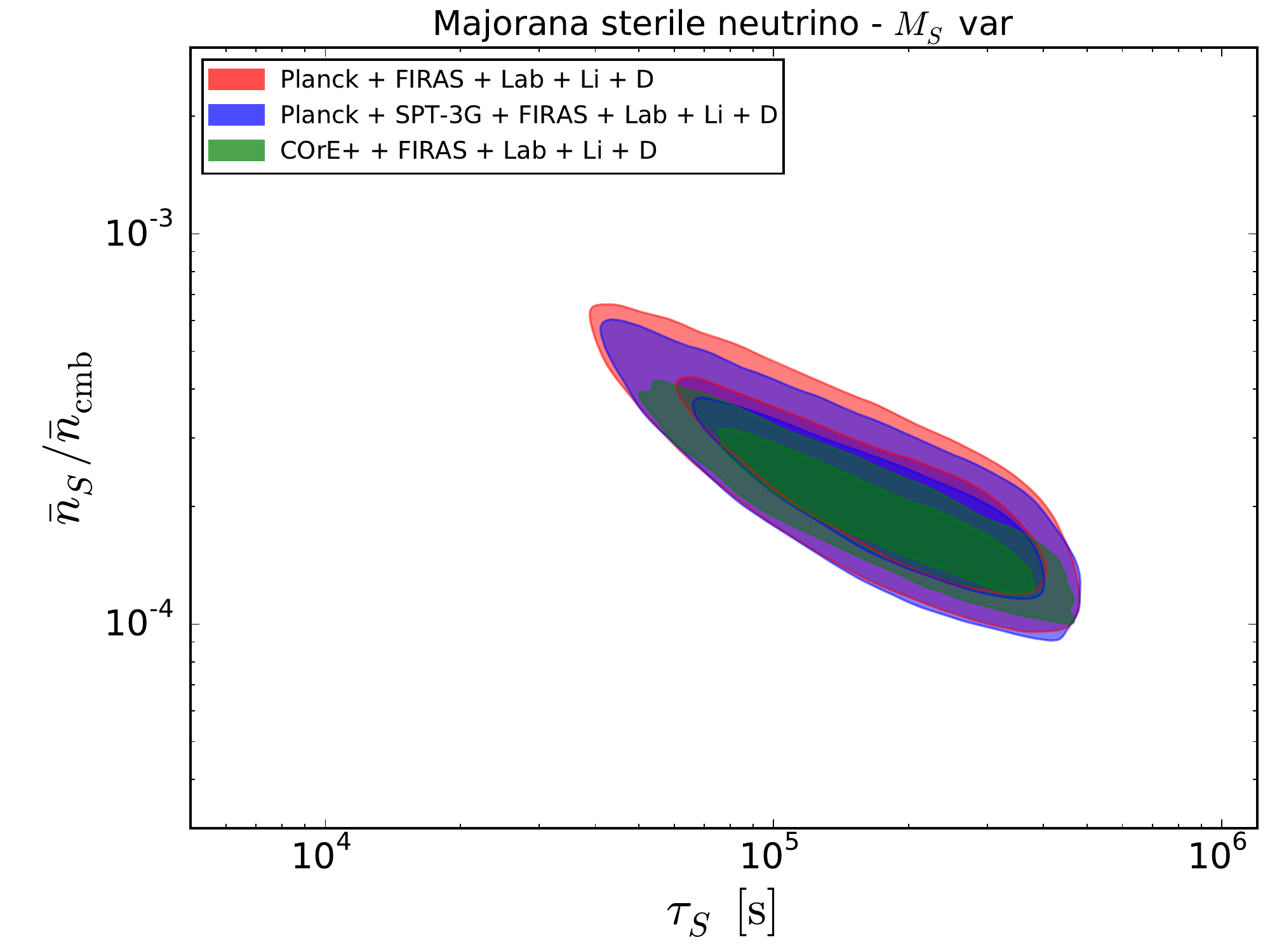}}
\caption{\footnotesize{Two dimensional distributions of sterile Majorana neutrino total decay time $\tau _S$ and comoving density $\bar{n}_S/\bar{n}_{\text{cmb}}$ for the different combinations of current and future datasets described in the text. On the left we report results without considering astrophysical observations of primordial abundances (blue and green contours). On the right we report results adding also astrophysical datasets. For visual reference, in both plots we show in red the most constraining results from current data. In the left panel the mass is fixed to $M_S = 4.4 \, \text{MeV}$. }}\label{fig:forecasts_CMB}
\end{figure}
\end{center}

In Figure \ref{fig:forecasts_CMB}, we show the two-dimensional probability contours in ($\tau _S$, $\bar{n}_S/\bar{n}_{\text{cmb}}$) plane. We compare forecasted results with the tightest constraints from current data, obtained considering also astrophysical observations. In the left panel of Figure \ref{fig:forecasts_CMB}, we have not included information from the astrophysical measurements. As a result, as discussed in the previous section, the lower bound on $\bar{n}_S/\bar{n}_{\text{cmb}}$ in Figure \ref{fig:forecasts_CMB} depends only on the capability of detecting $\Delta N_{\text{eff}}^{\text{cmb}}>0$. As we can see, the combination of current Planck data with forthcoming results from SPT-3G would not be able to reach the necessary sensitivity to detect the contribution of extra radiation ($\Delta N_{\text{eff}} ^{\text{cmb}}< 0.21$ at $95\%$ c.l.) coming from the sterile neutrino decay, sensitivity which would be instead achieved by the future COrE+ mission ($\Delta N_{\text{eff}}^{\text{cmb}} = 0.268\pm0.093$ at $95\%$ c.l.). However, if we consider results from CMB data alone, the expected improvement of CMB anisotropies measurements would not lead to more stringent constraints with respect to results from current data. In order to get an improvement, we need to combine future CMB datasets with direct astrophysical observations of primordial abundances, as we can see from the right panel of Figure. \ref{fig:forecasts_CMB}.

\begin{figure}[H]
\centering
\includegraphics[scale=0.5]{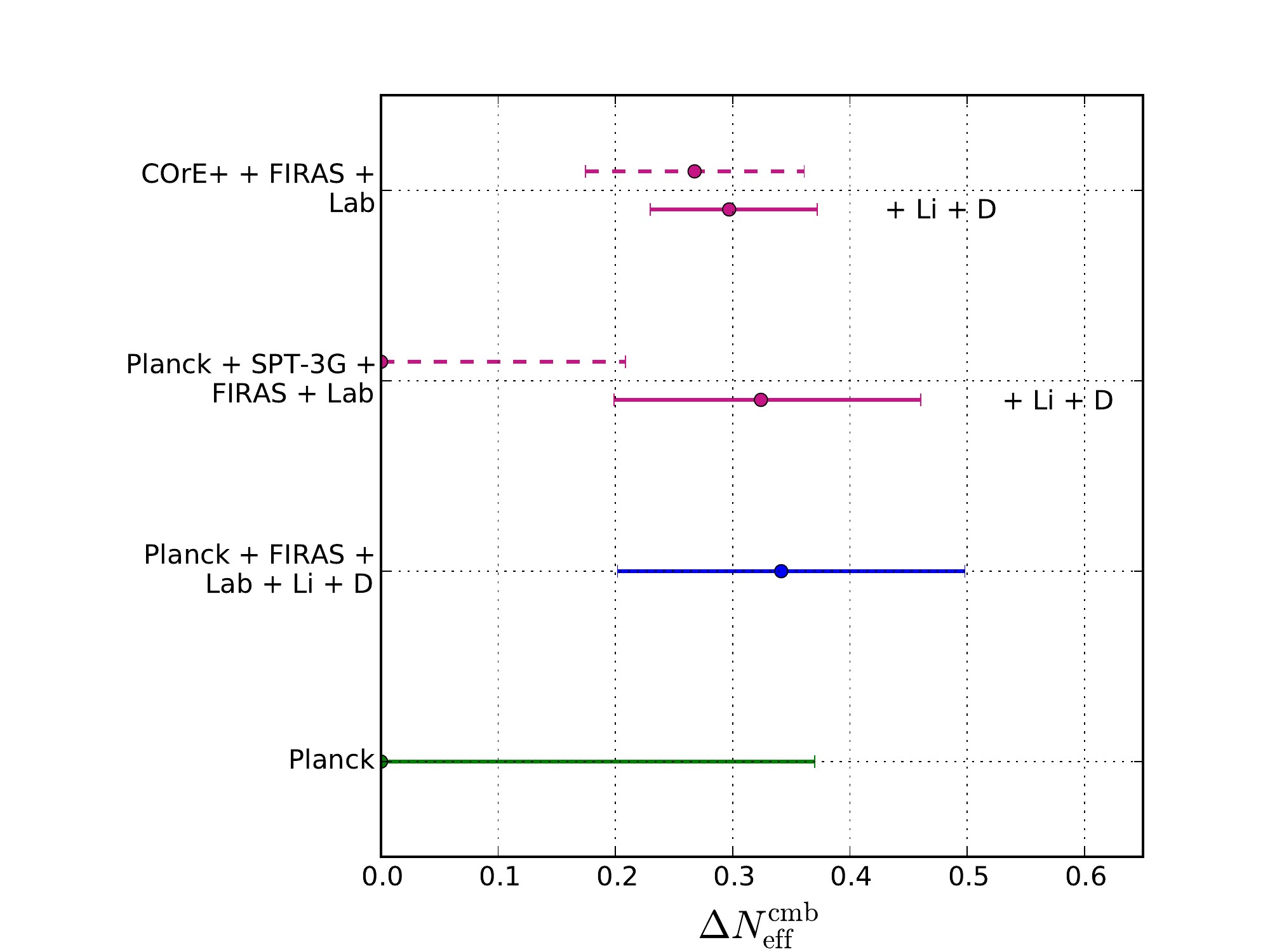}
\caption{\footnotesize{$95 \%$ c.l. of $\Delta N_{\text{eff}}^{\text{cmb}}$ for different dataset combinations. The dashed lines reproduce the limits for the CMB forecasts analysis without astrophysical observations, while the corresponding solid lines report results with the addition of them.}}
\label{fig:delta_Neff}
\end{figure}

In Figure \ref{fig:delta_Neff}, we report the $95\%$ c.l. for $\Delta N_{\text{eff}}^{\text{cmb}}$, in the model under consideration, for different combinations of datasets, both current and future. For comparison we also show the constraint on  $\Delta N_{\text{eff}}^{\text{cmb}}$ in a simple one-parameter extension of $\Lambda$CDM, from Planck data. The dashed lines refer to results obtained without considering information from astrophysical observations. As anticipated, the Planck + SPT-3G + FIRAS + Lab will not be able to constrain $\Delta N_{\text{eff}} ^{\text{cmb}} \neq 0$ if astrophysical measurements are not considered. In contrast, even if direct measurements of \Li\ and D abundances are not taken into account, the future COrE+ mission complemented with FIRAS and laboratory results will reach, in principle, the required sensitivity for detecting $\Delta N_{\text{eff}} ^{\text{cmb}} \neq 0$. The addition of astrophysical measurements to Planck + SPT-3G + FIRAS + Lab and COrE + FIRAS + Lab would lead to limits on $\Delta N_{\text{eff}} ^{\text{cmb}}$ slightly tighter and nearly a factor-of-two tighter than the Planck + FIRAS + Lab + Li + D case, respectively. Finally, we note that the lower limits for the Planck + FIRAS + Lab + Li + D, Planck + SPT-3G + FIRAS + Lab + Li + D and COrE+ + FIRAS + Lab + Li + D dataset combinations correspond to the lower bounds on $\bar{n}_S/\bar{n}_{\text{cmb}}$ that can be seen, e.g., in the right panel of Figure \ref{fig:forecasts_CMB}.

\begin{center}
\begin{figure}[H]
\subfigure[{}]%
{\includegraphics[scale=0.38]{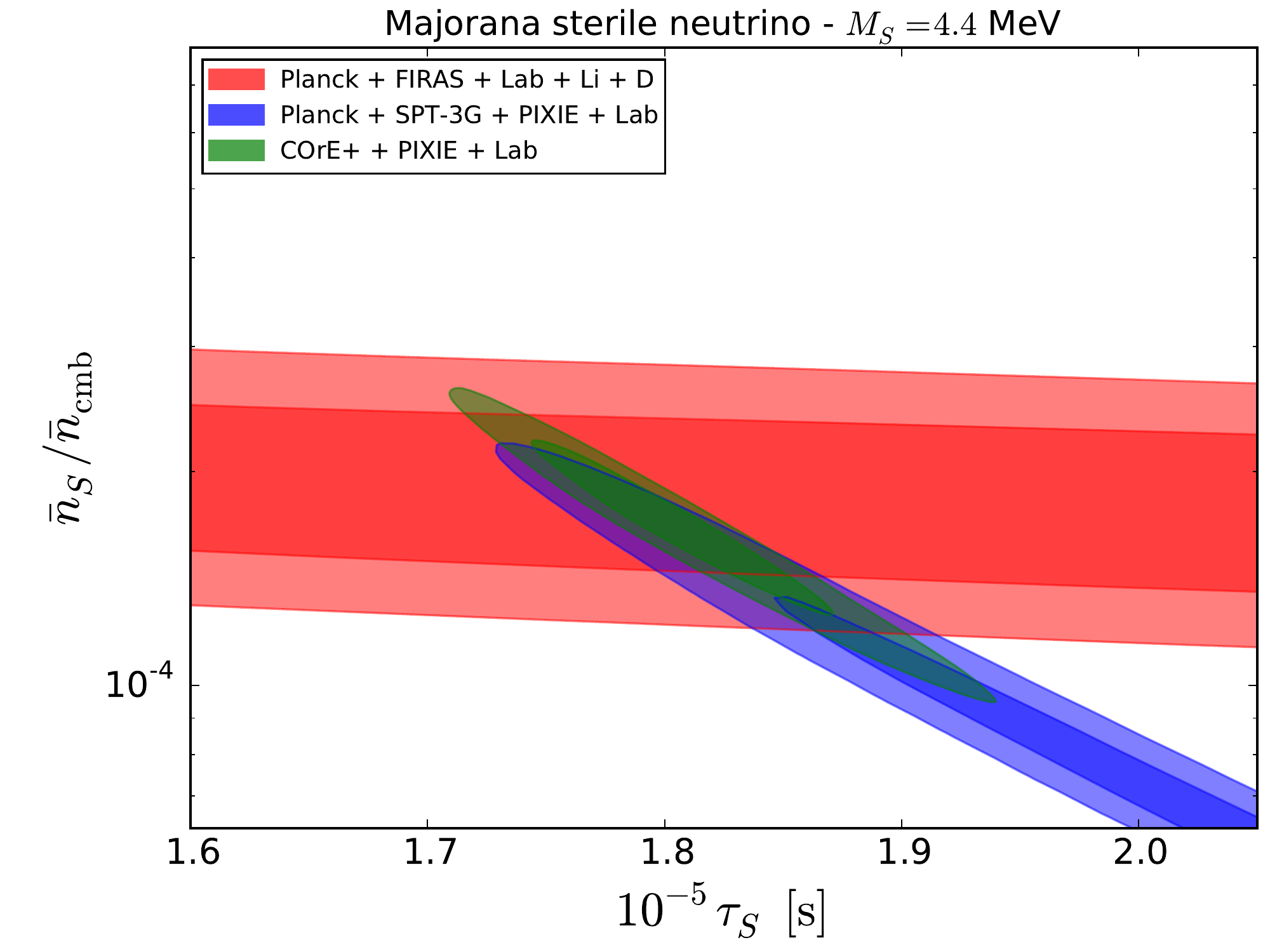}}
\subfigure[{}]%
{\includegraphics[scale=0.38]{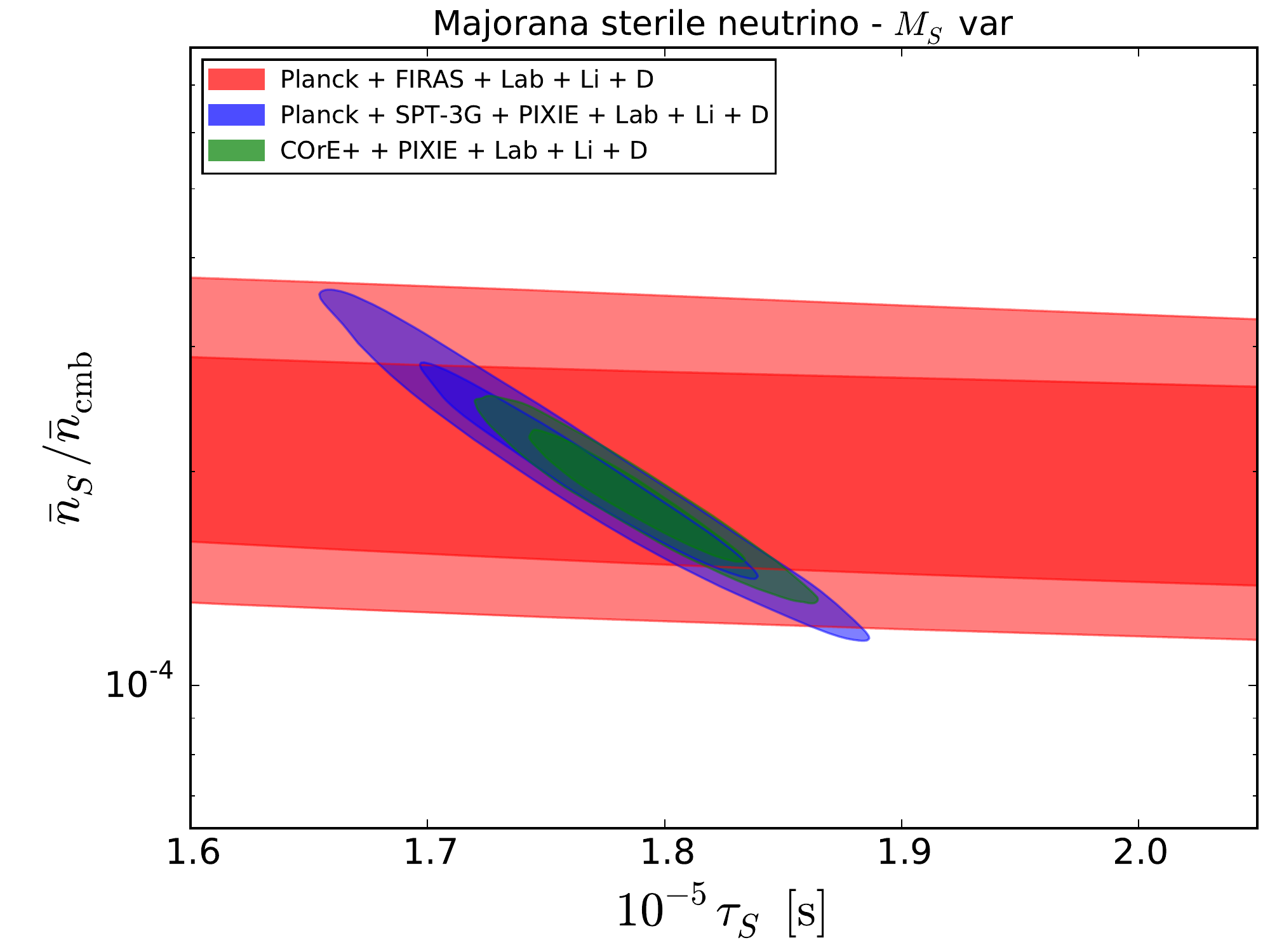}}
\caption{\footnotesize{Two dimensional distributions of sterile Majorana neutrino total decay time $\tau _S$ and comoving density $\bar{n}_S/\bar{n}_{\text{cmb}}$ for the different combinations of future datasets described in the text. On the left we report results without considering astrophysical observations of primordial abundances. On the right we report results adding also astrophysical datasets (blue and green contours). For visual reference, in both plots we show in red the most constraining results from current data. In the left panel the mass is fixed to $M_S = 4.4 \, \text{MeV}$. Please note the scale difference
in $\tau_S$ with respect to Figure \ref{fig:forecasts_CMB} and the linear scale on the horizontal axis.}}\label{fig:forecasts_CMB_SD}
\end{figure}
\end{center}

In Figure \ref{fig:forecasts_CMB_SD} we replace FIRAS results with predictions for a future detection of non-vanishing $\mu$-type spectral distortions with the PIXIE mission. Given the fiducial values of the parameters\footnote{We ignore distortions naturally produced in a $\Lambda$CDM model \cite{Chluba:2016bvg}.}, we assume a measurement $\mu = (3.0 \pm 0.1) \cdot 10^{-7}$. Also in this case, we compare forecasted results with the tightest constraints from current data, obtained considering also astrophysical observations. 
A future detection of CMB spectral distortions will produce a remarkable reduction in the parameter space allowed by the model. Even if we do not consider astrophysical observations (left panel), the COrE+ + PIXIE + Lab dataset combination leads to extremely tight constraints. In fact, a $\mu$-distortion detection mainly impacts the sterile neutrino lifetime, as discussed in the previous section, reducing the allowed range and producing stronger constraints on the energy parameter for the electromagnetic decay channels. The latter effect leads in turn to better bounds on the initial comoving density (which is clear if one remembers the definition of the energy parameter, see Eq. \ref{eq:zeta_def}). The combination of these effects along with the increased sensitivity in detecting $\Delta N_{\text{eff}}^{\text{cmb}}$ (from COrE+ mission) noticeably improves the constraints on the parameters that characterize this model. 

\begin{figure}[H]
\centering
\includegraphics[scale=0.5]{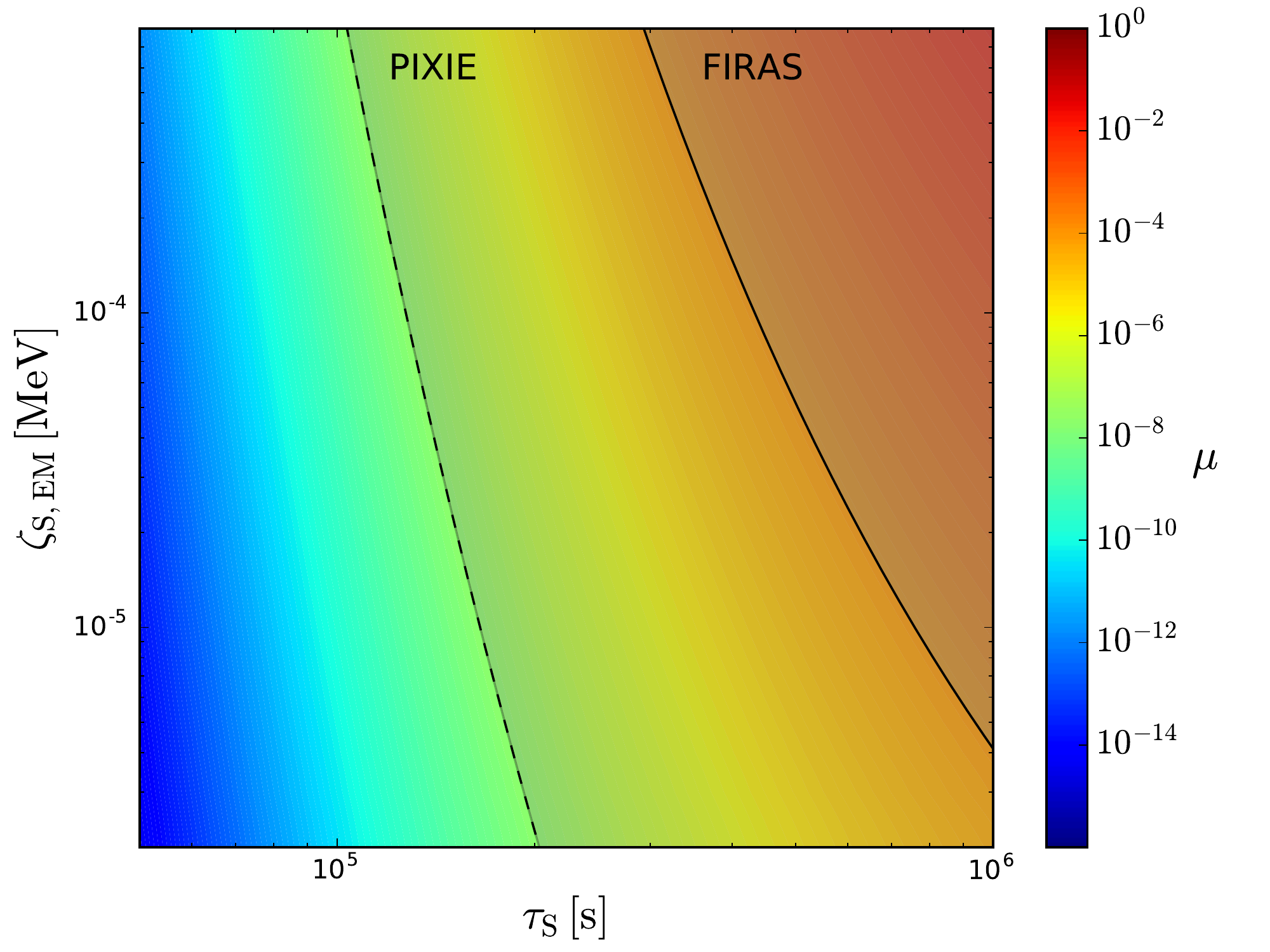}
\caption{\footnotesize{Contour plot of $\mu$-type distortions as function of the energy parameter for electromagnetic decay channels $\zeta_{S,\text{EM}}$ and the total decaytime $\tau _S$. We show the regions excluded by FIRAS \cite{Fixsen:1996nj} and PIXIE \cite{Kogut:2011xw} (in case of no detection). For our best-fit  $\zeta_{S,\text{EM}} \simeq 17.4\, \zeta_{S,\gamma}$.}}
\label{fig:SD}
\end{figure}

In Figure \ref{fig:SD}, we show the expected amount of spectral distortions, given the parameters of the model, together with the upper bounds from FIRAS and PIXIE (in case of no detection).

\section{Discussion and Conclusions}\label{sec:conclusions}

In this paper we have investigated the possibility to solve the cosmological lithium problem with the decay of a heavy sterile neutrino, in light of
the most precise data available to date both from cosmology and laboratory experiments. We have found that, for a mass of the sterile neutrino in the MeV range, the energy injected through the radiative decay of the heavy neutrino is able to photo-dissociate $^7 \text{Be}$, preventing it to be converted into $^7 \text{Li}$. By considering the impact that decay products from the different decay channels have on cosmological observables, we can confirm that this model is definitely allowed by current cosmological data, as well as in agreement with laboratory bounds on the sterile-active neutrino mixing angles. We have extended the analysis of Ref. \cite{Poulin:2015woa}, by performing a fully consistent statistical analysis in a Bayesian framework, to show how the combination of CMB measurements (temperature and polarization anisotropies and spectral distortions), direct astrophysical observations of primordial nuclei and laboratory bounds on neutrino mixing angles are able to put strong constraints on the relevant parameters of the model. We note that our results strongly disfavour the possibility of a sterile neutrino that mostly mixes with $\nu_e$,
like in the model considered in Ref. \cite{Ishida:2014wqa}.

We have also performed forecasts for constraints from future CMB anisotropies and spectral distortions data. The improved sensitivity of future experiments will result in more stringent constraints. In particular, the realization of the model that provides the best joint fit to current CMB, abundances and laboratory data,
predicts an amount of $\mu$-type spectral distortions $\mu \simeq 3 \cdot 10^{-7}$, well within the reach of the PIXIE satellite.
A detection of $\mu$-type spectral distortions from the PIXIE satellite would provide strong constraints on the lifetime $\tau _S$ of the heavy neutrino. Moreover, the model
also predicts a value of the effective number of relativistic species $N_\mathrm{eff} \simeq 3.3 $, that can be distinguished from the standard value of 3.046
by a space mission devoted to the study of CMB anisotropies like COrE+.

We have not considered here the implications that improved measurements of the sterile-active neutrino mixing angles could have in the future. Notice that current constraints (in combination with cosmological and astrophysical data) already reduce the region of parameter space allowed for this model, pointing to a sterile neutrino that mostly mixes with the $\tau$ neutrino. The preferred values of the $\theta_\tau$ mixing angle lie not much below the upper bounds currently available from laboratory experiments. 
Improving the sensitivity of laboratory experiments to $\theta^2_\tau$ by an order of magnitude would lead to a detection of a non-zero mixing angle (and thus to the discovery of the existence of a fourth neutrino eigenstate) if the model considered here is the actual explanation to the lithium discrepancy. Conversely,
more sensitive laboratory searches would exclude the model by finding a tighter upper bound on the mixing.

We also mention that sterile neutrinos could be produced in a supernova core, so that their mass and mixing angles can be constrained by the observations of core-collapse SuperNovae (SN). In fact, observations of SN1987A can be used to this purpose, as done in \cite{Gelmini:2008fq}  for the mass range of interest here.
These limits are much more stringent than the ones obtained from direct laboratory experiments, and, if taken at face value, would rule out the heavy neutrino
solution to the lithium problem. However, there are still non negligible uncertainties on these limits, due to the incomplete knowledge of the physics controlling SN explosion. For this reason, we decide to regard them as only disfavouring the model studied in this paper.  

We also should note that in our analysis we have not assumed any specific mechanism for the production of sterile neutrinos in the early Universe. This is the reason
why we have treated the initial comoving density $\bar n_S$ of the sterile neutrino as a quantity independent from the parameters of the underlying particle physics model (in this specific case, the mass and the mixing angles). However, after having derived our constraints, we can ask whether the preferred solution for the lithium problem that involves a heavy neutrino can be realized in a specific framework for sterile neutrino production. We find that, for the values of the mixing angles allowed by the model, the required primordial density cannot be produced through the Dodelson-Widrow mechanism, at least in the standard cosmological scenario. In fact, our analysis suggests an abundance of  $\bar{n}_S/\bar{n}_{\rm cmb} \sim 10^{-3} \div 10^{-4}$. The production of a MeV-mass neutrino mostly happens when $T=T_\mathrm{max} \sim 1\,\GeV$, and for $\Theta^2 \sim 10^{-4}$, as required to solve the lithium problem, the sterile would still be coupled to the plasma at that time, thus leading to a thermal abundance. This is clearly at variance with the requirement that $\bar{n}_s \sim 10^{-3} \div 10^{-4}\,\bar{n}_\mathrm{cmb}$. In fact, production of a sterile neutrino with $M_S \sim \MeV$ and $\Theta^2 \sim 10^{-4}$ through mixing is excluded by arguments based on the expansion history of the Universe, see e.g. Ref. \cite{Vincent:2014rja}.

A possible way to circumvent this advocates a low-reheating temperature $T_{\text{RH}}$, as detailed in \cite{Gelmini:2008fq}. If $T_{\text{RH}} < T_\mathrm{max}$, the abundance of sterile neutrinos is suppressed. Following Eq. (11) and Eq. (12) in \cite{Gelmini:2008fq}, we find that for $2.8 \MeV \lesssim T_{\text{RH}} \lesssim 6.1\, \text{MeV}$ the required abundance is generated. From \cite{Gelmini:2008fq}, we can see that the recovered reheating temperature is much lower than the temperature corresponding to the maximum production rate ($T_{\text{RH}} \ll T_{\text{max}} \simeq 2 \, \text{GeV} $). 

In addition, the reheating temperature can be constrained by employing CMB and BBN measurements. In particular, in Ref. \cite{deSalas:2015glj}, lower limits of $T_{\text{RH}}>  4.7 \,\text{MeV}$ and $T_{\text{RH}}> 4.3 \, \text{MeV}$ from CMB and BBN respectively, have been derived. By comparing the former results with the values of $T_{\text{RH}}$ recovered from our analysis, we note that there is a range, though small, of overlapping values, suggesting that the model presented here, along with the suggested production mechanism, is physically permitted. A low reheating temperature affects also the number of extra relativistic degrees of freedom, resulting in $N_{\text{eff}}^{\text{cmb}}< 3.046$. We have seen that neutrino production accompanies the sterile neutrino decay, leading to $\Delta N_{\text{eff}} ^{\text{cmb}}> 0$. Thus, the two effects of a low-reheating temperature and extra-production of neutrinos could possibly balance, so we argue that a fully consistent analysis of CMB data
in the framework of a scenario with a heavy neutrino and a low reheating temperature could allow lower values of $T_\text{RH}$. The same effect would not work 
when constraining $T_\text{RH}$ with element abundances, these being evaluated before sterile neutrinos decay. As a result, we can assume $T_\text{RH} > 4.3 \, \text{MeV}$ as the more conservative lower limit.

In view of the above, we can safely confirm that the model investigated in this analysis is in agreement with a production mechanism induced by a low reheating temperature scenario.

In our analysis we have focused on a specific particle physics scenario, involving a heavy, mostly sterile neutrino. Other scenarios involving
the radiative decay of a MeV-mass particle are however possible. For example, viable candidates could be the Majoron in a general seesaw model (where the decay
to photon is induced by the coupling to a Higgs triplet) \cite{Chikashige:1980ui,Schechter:1980gr,Bazzocchi:2008fh},
or a light Gravitino in supergravity models. Also, different implementations of the sterile neutrino model, less minimal than the one studied here,
can be considered, like e.g. in gauge extensions of the standard model \cite{Bezrukov:2009th}.

\begin{figure}[H]
\centering
\includegraphics[scale=0.5]{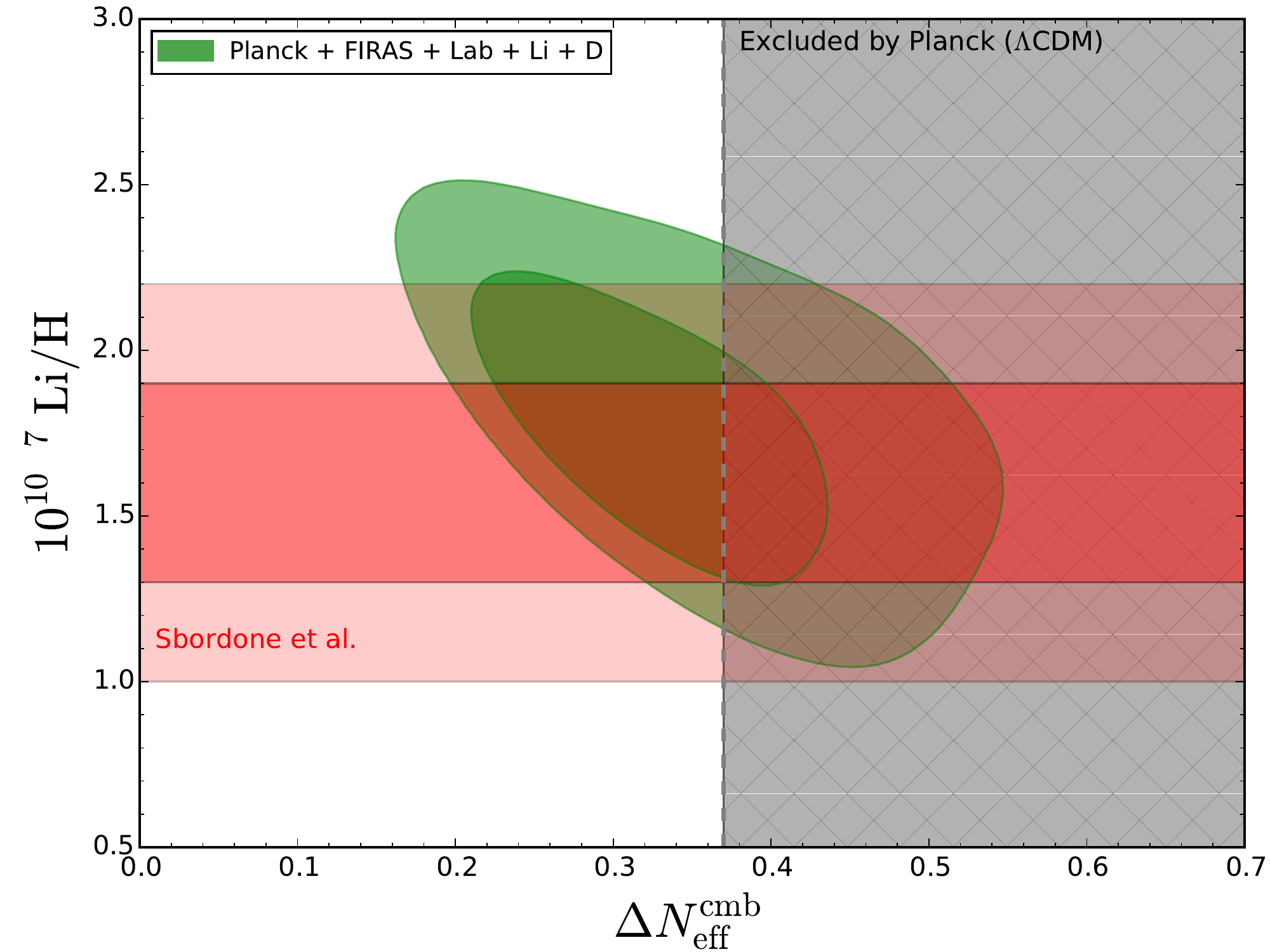}
\caption{\footnotesize{Two dimensional posterior distributions for lithium abundance and $\Delta N_{\text{eff}}$ parameter. The red horizontal band is the $^7 \text{Li}/\text{H}$ astrophysical measurement \cite{Sbordone:2010zi} ($68\% $ and $95\%$ c.l.) while the hatched region is excluded from Planck 2015 constraints on $N_{\text{eff}}$ \cite{Ade:2015xua}.}}
\label{fig:Li-Neff}
\end{figure}

Finally, we notice that the current and future upper limits on $\Delta N_{\text{eff}}^{\text{cmb}}$ as determined by CMB and BBN measurements can be translated into a lower limit of \Li\ abundance (see Figure \ref{fig:Li-Neff}). If future astrophysical measurements would provide constraints in agreement with an even lower abundance of \Li\ with respect to the lower limit predicted by the above mechanism, this would automatically rule out the model. On the other hand, even too low a value of $\Delta N_{\text{eff}}^{\text{cmb}}$ ($\lesssim 0.1$) would make
the proposed scenario hardly compatible with current Lithium astrophysical measurement.

\section*{Acknowledgments}
ML acknowledges support from ASI through ASI/INAF Agreement I/072/09/0 for the Planck LFI Activity of Phase E2. MG was partly supported by the grant ``Avvio alla ricerca''  for young researchers by ``Sapienza'' university and is supported by the Vetenskapsr\aa det (Swedish Research Council). 


\pagebreak
\appendix

\section{Heavy neutrino decays to charged leptons \label{app:decay_rates}}

Decay rates of heavy neutrinos to charged leptons pairs $\ell_\beta^+\ell_\beta^-$ accompanied by an active neutrino $\nu_\alpha$ are given in Ref. \cite{Gorbunov:2007ak}:

\begin{align}
\Gamma_{\nu_S \to \nu_\alpha \ell_{\beta\ne\alpha}^+ \ell_{\beta\ne\alpha}^-} = 
\frac{G_F^2 M_S^5}{192 \pi^3} \cdot \theta^2 _{\alpha} \cdot \Bigg\{ C_1\bigg[\Big(1-14 x_\ell^2 - 2 x_\ell^4 -12 x_\ell^6 \Big)\sqrt{1-4 x_\ell^2} + 12 x_\ell^4 \Big(x_\ell^4-1\Big) L\bigg] + \nonumber \\
+ 4 C_2 \bigg[ x_\ell^2 \Big( 2+ 10 x_\ell^2 -12 x_\ell^4 \Big) \sqrt{1-4 x_\ell^2}  + 6 x_\ell^4 \Big(1 - 2 x_\ell^2 + 2 x_\ell^4\Big) L \bigg] \Bigg\} \\
\Gamma_{\nu_S \to \nu_\alpha \ell_{\beta=\alpha}^+ \ell_{\beta=\alpha}^-} = 
\frac{G_F^2 M_S^5}{192 \pi^3} \cdot \theta^2 _{\alpha} \cdot \Bigg\{ C_3\bigg[\Big(1-14 x_\ell^2 - 2 x_\ell^4 -12 x_\ell^6 \Big)\sqrt{1-4 x_\ell^2} + 12 x_\ell^4 \Big(x_\ell^4-1\Big) L\bigg] + \nonumber \\
+ 4 C_4 \bigg[ x_\ell^2 \Big( 2+ 10 x_\ell^2 -12 x_\ell^4 \Big) \sqrt{1-4 x_\ell^2}  + 6 x_\ell^4 \Big(1 - 2 x_\ell^2 + 2 x_\ell^4\Big) L \bigg] \Bigg\}\label{eq:decay2leptons}
\end{align}
where $x_\ell \equiv m_\ell/M_S$, the $C$'s are functions of the weak angle $\theta_W$,
\begin{align}
C_1 \equiv \frac{1}{4}\left( 1 - 4 \sin^2 \theta_W + 8 \sin^4\theta_W \right); \qquad C_2 \equiv \frac{1}{2}\sin^2\theta_W\left(2\sin^2\theta_W-1\right) \, ,\\
C_3 \equiv \frac{1}{4}\left( 1 + 4 \sin^2 \theta_W + 8 \sin^4\theta_W \right); \qquad C_4 \equiv \frac{1}{2}\sin^2\theta_W\left(2\sin^2\theta_W+1\right) \, ,
\end{align}
and 
\begin{equation}
L\equiv \ln\left[\frac{1-3x_\ell^2-\left(1-x_\ell^2\right)\sqrt{1-4x_\ell^2}}{x_\ell^2\left(1+\sqrt{1-4x_\ell^2}\right)}\right]
\end{equation}
Specifying to the case of decay to electrons, and summing over the flavour of the neutrino in the final state, we obtain the second of Eqs.~\eqref{eq:dec_ch}, having defined:
\begin{align}
f(x) \equiv C_3\bigg[\Big(1-14 x^2 - 2 x^4 -12 x^6 \Big)\sqrt{1-4 x^2} + 12 x^4 \Big(x^4-1\Big) L\bigg] + \nonumber \\
+ 4 C_4 \bigg[ x^2 \Big( 2+ 10 x^2 -12 x^4 \Big) \sqrt{1-4 x^2}  + 6 x^4 \Big(1 - 2 x^2 + 2 x^4\Big) L \bigg] \, , \label{eq:fx}\\
g(x) \equiv C_1\bigg[\Big(1-14 x^2 - 2 x^4 -12 x^6 \Big)\sqrt{1-4 x^2} + 12 x^4 \Big(x^4-1\Big) L\bigg] + \nonumber \\
+ 4 C_2 \bigg[ x^2 \Big( 2+ 10 x^2 -12 x^4 \Big) \sqrt{1-4 x^2}  + 6 x^4 \Big(1 - 2 x^2 + 2 x^4\Big) L \bigg] \, . \label{eq:gx}
\end{align}
In our calculations, we use $\sin^2 \theta_W = 0.23$, so that $C_1 = 0.1258$, $C_2 = - 0.0621$, $C_3 = 0.5858$, and $C_4 = 0.1679$.

\section{Interaction rates}\label{app:int_rates}
In this section we report the computational details to evaluate the total interaction rate between injected photons and primordial plasma. As shown in section \ref{sec:NtBBN}, this quantity is the sum of different contributions, relevant in the energy range that we are considering. In particular we consider Compton scattering over thermal electrons ($\Gamma_{\text{Com}}$), photon scattering ($\Gamma_{\text{PS}}$), pair production over nuclei ($\Gamma_{\text{PPn}}$) and pair production over photons ($\Gamma _{\text{PP}\gamma}$).

We start considering Compton scattering over thermal electrons. Following \cite{1989ApJ...344..551Z} we derive the interaction rate, starting from:
\begin{equation}
\dfrac{d\tau}{dl} = \tau_0 (1+z)^3 f(x) = \dfrac{\Gamma_{\text{Com}}}{H_0}
\end{equation}
where $x=E_{\gamma}/(m_e c^2)$. We define
\begin{eqnarray}
\tau_0 &=& \sigma _T n_e^0 \dfrac{c}{H_0} = \dfrac{8 \pi r_e^2}{3}n_e^0 \dfrac{c}{H_0}  \label{eq:def1}\\
 n_e &=& n_b \left( \dfrac{1+2f_{\text{He}}}{1+4f_{\text{He}}} \right) = \dfrac{\Omega_bh^2 \rho _{\text{cr}}}{m_p h^2} \left( \dfrac{1+2f_{\text{He}}}{1+4f_{\text{He}}} \right)   \label{eq:def2}\\
  f(x) &=& \dfrac{3}{8x} \left[ \left( 1-\dfrac{2}{x} -\dfrac{2}{x^2} \right)\text{ln}(1+2x) +\dfrac{4}{x}+ \dfrac{2x(1+x)}{(1+2x)^2}  \right] \label{eq:def3}
\end{eqnarray}
such that
\begin{equation}
\Gamma _{\text{Com}} = n_b\, c\, \sigma _T \left( \dfrac{1+2f_{\text{He}}}{1+4f_{\text{He}}} \right) (1+z)^3 f(x)
\end{equation}

where we have used the cross section for Thomson scattering $\sigma_T$, the critical density of the Universe $\rho _{\text{cr}}$, the electron and baryon number density $n_e$ and $n_b$, the baryon density parameter $\Omega_b$ and $f_{\text{He}}\simeq Y_{\text{P}}/(4(1-Y_{\text{P}}))$.

We then consider Photon scattering, again following \cite{1989ApJ...344..551Z}. We start from 
\begin{equation}
\dfrac{d\tau}{dl} = \tau_0 (1+z)^6 x^3 = \dfrac{\Gamma_{\text{PS}}}{H_0}
\end{equation}
and using the same quantitites defined above in Eq. (\ref{eq:def1}) and (\ref{eq:def2}) we obtain
\begin{equation}\label{eq:Gamma_PS}
\Gamma_{\text{PS}} =  n_b\, c\, \sigma _T  \left( \dfrac{1+2f_{\text{He}}}{1+4f_{\text{He}}} \right) (1+z)^6 x ^3
\end{equation}

For the Pair production over nuclei we follow what is reported in \cite{1995ApJ...452..506K}. We consider interactions of photons with both nuclei of $\text{H}$ and $^4 \text{He}$
\begin{equation}\label{eq:Gamma_PPn}
\Gamma_{\text{PPn}} =
\left\{
\begin{array}{rl}
n_b\, c \left[ \sigma_1^{\text{H}}(x) \dfrac{2+2f_{\text{He}}}{1+4f_{\text{He}}}+ \sigma_1^{\text{He}}(x) \dfrac{f_{\text{He}}}{1+4f_{\text{He}}}\right] (1+z)^3& \mbox{if } x <4 \\
 & \\
n_b\, c \left[ \sigma_2^{\text{H}}(x) \dfrac{2+2f_{\text{He}}}{1+4f_{\text{He}}}+ \sigma_2^{\text{He}}(x) \dfrac{f_{\text{He}}}{1+4f_{\text{He}}}\right] (1+z)^3& \mbox{if } x \geq 4
\end{array}
\right.
\end{equation}
where $\sigma_1$ and $\sigma_2$ are taken from \cite{1995ApJ...452..506K} eqs. (36) and (38), $n_b$, $f_{\text{He}}$ and $x$ are defined as above.\\

Finally we evaluate the interaction rate for Pair production over photons, again taken from \cite{1989ApJ...344..551Z}.
\begin{eqnarray}\label{eq:Gamma_PPg}
\Gamma _{\text{PP}\gamma} &=& n_b\, c\, \sigma _T \left( \dfrac{1+2f_{\text{He}}}{1+4f_{\text{He}}} \right) \left[\dfrac{ 2\sqrt{\pi}}{\sqrt{y} \, \text{e}^{1/y}} \left( 1+\dfrac{9}{4}y \right) \right] (1+z)^3 \\
y &=& x \dfrac{k_{\text{B}}T_{\text{CMB},0}}{m_ec^2}(1+z)  \ll 1 \nonumber
\end{eqnarray}
and $x$ is defined as above.\\

\begin{figure}[H]
\centering
\includegraphics[scale=0.5]{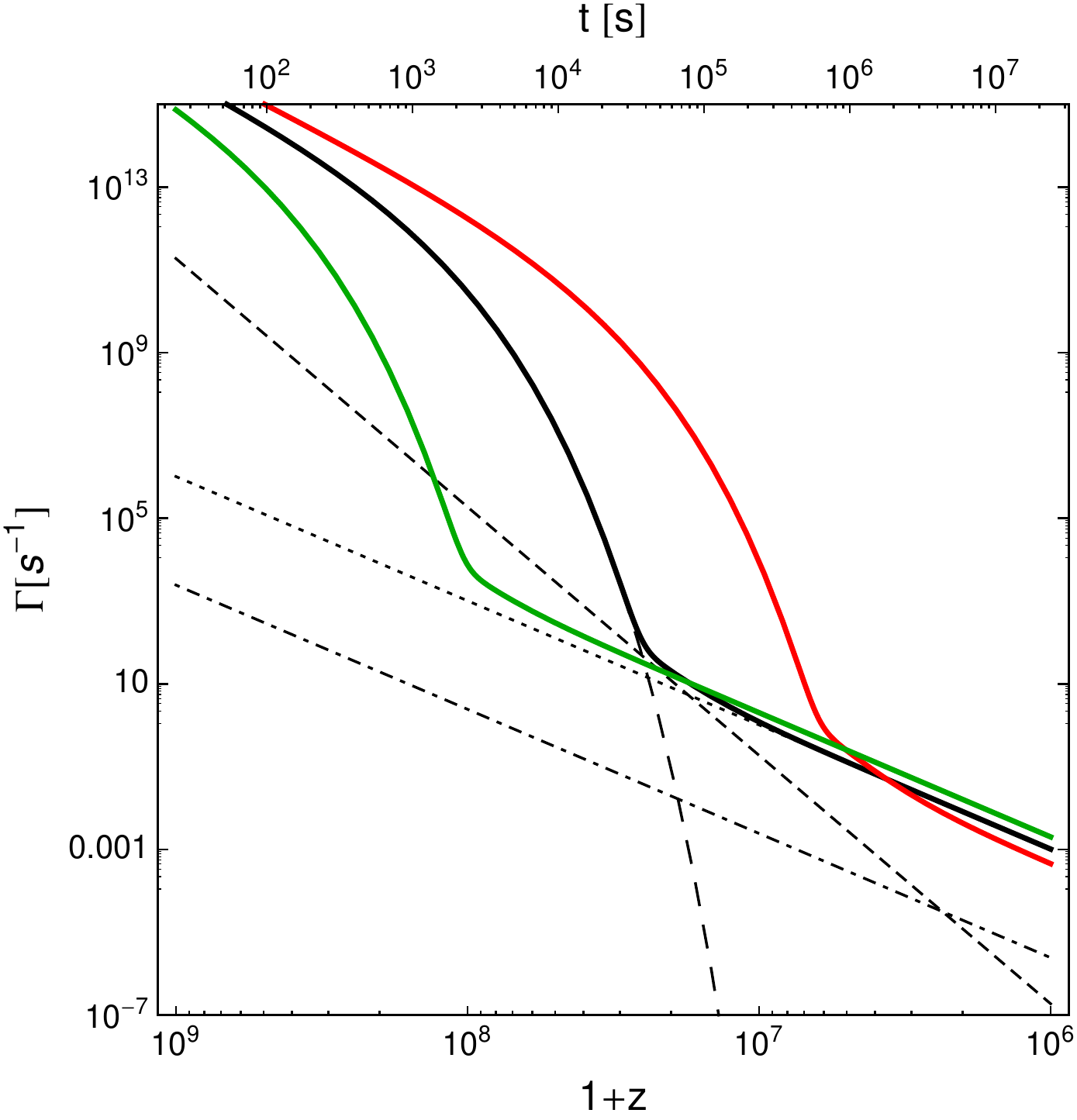}
\caption{\footnotesize{We show the total interaction rates for different energy ($E_0$) values: 0.5 (green), 2 (black), 8(red) MeV. We draw also the different contributions from the single interactions for the 2-MeV case: Compton scattering (dotted line), photon scattering (short dashed), pair production over nuclei (dot dashed) and pair production over photons (long dashed).}}
\label{fig:int_rates}
\end{figure}

In Figure \ref{fig:int_rates} we report the total interaction rates for different values of the injected energy $E_0 = 0.5 \, \text{MeV},\,2 \, \text{MeV},\, 8\, \text{MeV}$. In particular, for $E_0 = 2 \, \text{MeV}$ we report also the single contributions for the different interactions.

\begin{thebibliography}{99}

\bibitem{Cyburt:2015mya}
  R.~H.~Cyburt, B.~D.~Fields, K.~A.~Olive and T.~H.~Yeh,
  arXiv:1505.01076 [astro-ph.CO].

\bibitem{Cooke:2013cba}
  R.~Cooke, M.~Pettini, R.~A.~Jorgenson, M.~T.~Murphy and C.~C.~Steidel,
  Astrophys.\ J.\  {\bf 781} (2014) 1,  31
 
\bibitem{Aver:2015iza}
  E.~Aver, K.~A.~Olive and E.~D.~Skillman,
  JCAP {\bf 1507} (2015) 07,  011

\bibitem{Ade:2015xua}
  P.~A.~R.~Ade {\it et al.} [Planck Collaboration],
  arXiv:1502.01589 [astro-ph.CO].

\bibitem{Adam:2015rua}
  R.~Adam {\it et al.} [Planck Collaboration],
  arXiv:1502.01582 [astro-ph.CO].

\bibitem{Fields:2011zzb}
  B.~D.~Fields,
  Ann.\ Rev.\ Nucl.\ Part.\ Sci.\  {\bf 61} (2011) 47

\bibitem{Steigman:2007xt}
  G.~Steigman,
  Ann.\ Rev.\ Nucl.\ Part.\ Sci.\  {\bf 57} (2007) 463

\bibitem{Fields:2014uja}
  B.~D.~Fields, P.~Molaro and S.~Sarkar,
  Chin.\ Phys.\ C {\bf 38} (2014)  
  
 \bibitem{Sbordone:2010zi}
  L.~Sbordone {\it et al.},
  Astron.\ Astrophys.\  {\bf 522} (2010) A26

\bibitem{Spite:1982dd}
  F.~Spite and M.~Spite,
  Astron.\ Astrophys.\  {\bf 115} (1982) 357.
  
 \bibitem{Jedamzik:2004er}
  K.~Jedamzik,
  Phys.\ Rev.\ D {\bf 70} (2004) 063524

\bibitem{Jedamzik:2005dh}
  K.~Jedamzik, K.~Y.~Choi, L.~Roszkowski and R.~Ruiz de Austri,
  JCAP {\bf 0607} (2006) 007

\bibitem{Jedamzik:2007cp}
  K.~Jedamzik,
  Phys.\ Rev.\ D {\bf 77} (2008) 063524

\bibitem{Kusakabe:2007fu}
  M.~Kusakabe, T.~Kajino, R.~N.~Boyd, T.~Yoshida and G.~J.~Mathews,
  Phys.\ Rev.\ D {\bf 76} (2007) 121302

\bibitem{Bailly:2008yy}
  S.~Bailly, K.~Jedamzik and G.~Moultaka,
  Phys.\ Rev.\ D {\bf 80} (2009) 063509
  
\bibitem{Cyburt:2010vz}
  R.~H.~Cyburt, J.~Ellis, B.~D.~Fields, F.~Luo, K.~A.~Olive and V.~C.~Spanos,
  JCAP {\bf 1010} (2010) 032

\bibitem{Kusakabe:2014ola}
  M.~Kusakabe, M.~K.~Cheoun and K.~S.~Kim,
  Phys.\ Rev.\ D {\bf 90} (2014) no.4,  045009
  
\bibitem{Ishida:2014wqa}
  H.~Ishida, M.~Kusakabe and H.~Okada,
  Phys.\ Rev.\ D {\bf 90} (2014) 8,  083519

  
\bibitem{Poulin:2015woa}
  V.~Poulin and P.~D.~Serpico,
  Phys.\ Rev.\ Lett.\  {\bf 114} (2015) 9,  091101
 
\bibitem{Goudelis:2015wpa} 
  A.~Goudelis, M.~Pospelov and J.~Pradler,
  Phys.\ Rev.\ Lett.\  {\bf 116}, no. 21, 211303 (2016)
  doi:10.1103/PhysRevLett.116.211303
  [arXiv:1510.08858 [hep-ph]].
 
\bibitem{Minkowski:1977sc} 
  P.~Minkowski,
  Phys.\ Lett.\ B {\bf 67}, 421 (1977).
  
\bibitem{Yanagida:1979as} 
  T.~Yanagida,
  Conf.\ Proc.\ C {\bf 7902131}, 95 (1979).
  
\bibitem{Mohapatra:1979ia} 
  R.~N.~Mohapatra and G.~Senjanovic,
  Phys.\ Rev.\ Lett.\  {\bf 44}, 912 (1980).
  
\bibitem{GellMann:1980vs} 
  M.~Gell-Mann, P.~Ramond and R.~Slansky,
  Conf.\ Proc.\ C {\bf 790927}, 315 (1979)
  [arXiv:1306.4669 [hep-th]].

\bibitem{Schechter:1980gr} 
  J.~Schechter and J.~W.~F.~Valle,
  Phys.\ Rev.\ D {\bf 22}, 2227 (1980).
  
\bibitem{Gelmini:2008fq}
  G.~Gelmini, E.~Osoba, S.~Palomares-Ruiz and S.~Pascoli,
  JCAP {\bf 0810} (2008) 029
  
\bibitem{Denner:1992me} 
  A.~Denner, H.~Eck, O.~Hahn and J.~Kublbeck,
  Phys.\ Lett.\ B {\bf 291}, 278 (1992).

\bibitem{Johnson:1997cj} 
  L.~M.~Johnson, D.~W.~McKay and T.~Bolton,
  Phys.\ Rev.\ D {\bf 56}, 2970 (1997).
  
\bibitem{Lavoura:2003xp} 
  L.~Lavoura,
  Eur.\ Phys.\ J.\ C {\bf 29}, 191 (2003).
  
\bibitem{Gorbunov:2007ak} 
  D.~Gorbunov and M.~Shaposhnikov,
  JHEP {\bf 0710}, 015 (2007);
  Erratum: [JHEP {\bf 1311}, 101 (2013)].
  
\bibitem{Bezrukov:2009th} 
  F.~Bezrukov, H.~Hettmansperger and M.~Lindner,
  Phys.\ Rev.\ D {\bf 81}, 085032 (2010).
\bibitem{Cyburt:2002uv}
  R.~H.~Cyburt, J.~R.~Ellis, B.~D.~Fields and K.~A.~Olive,
  Phys.\ Rev.\ D {\bf 67} (2003) 103521

\bibitem{Chluba:2011hw}
  J.~Chluba and R.~A.~Sunyaev,
  Mon.\ Not.\ Roy.\ Astron.\ Soc.\  {\bf 419} (2012) 1294

\bibitem{Feng_entropy}
 J.L.~Feng, A.~Rajaraman and F.~Takayama,
 Phys.\ Rev.\ D {\bf 68} (2003) 063504

\bibitem{Aghanim:2015xee}
  N.~Aghanim {\it et al.} [Planck Collaboration],
  [arXiv:1507.02704 [astro-ph.CO]].
  
\bibitem{Agashe:2014kda}
  K.~A.~Olive {\it et al.} [Particle Data Group Collaboration],
  Chin.\ Phys.\ C {\bf 38} (2014) 090001.

\bibitem{Cooke:2016rky}
  R.~Cooke, M.~Pettini, K.~M.~Nollett and R.~Jorgenson,
  arXiv:1607.03900 [astro-ph.CO].

\bibitem{Adelberger:2010qa}
  E.~G.~Adelberger {\it et al.},
  Rev.\ Mod.\ Phys.\  {\bf 83} (2011) 195

\bibitem{Fixsen:1996nj}
  D.~J.~Fixsen, E.~S.~Cheng, J.~M.~Gales, J.~C.~Mather, R.~A.~Shafer and E.~L.~Wright,
  Astrophys.\ J.\  {\bf 473} (1996) 576

\bibitem{Helo:2011yg}
  J.~C.~Helo, S.~Kovalenko and I.~Schmidt,
  Phys.\ Rev.\ D {\bf 84} (2011) 053008

\bibitem{Benson:2014qhw}
  B.~A.~Benson {\it et al.} [SPT-3G Collaboration],
  Proc.\ SPIE Int.\ Soc.\ Opt.\ Eng.\  {\bf 9153} (2014) 91531P 
 
\bibitem{CORE} 
  http://conservancy.umn.edu/handle/11299/169642
  
\bibitem{Kogut:2011xw}
  A.~Kogut {\it et al.},
  JCAP {\bf 1107} (2011) 025  

\bibitem{Lewis:2002ah}
  A.~Lewis and S.~Bridle,
  Phys.\ Rev.\ D {\bf 66} (2002) 103511
  
\bibitem{Lewis:2013hha}
  A.~Lewis,
  Phys.\ Rev.\ D {\bf 87} (2013) 10,  103529
  
\bibitem[{Lewis et al.}(1999)]{Lewis:1999bs} 
  A.~Lewis, A.~Challinor and A.~Lasenby,
  Astrophys.\ J.\  {\bf 538}, 473 (2000).

    
\bibitem{Pisanti:2007hk}
  O.~Pisanti, A.~Cirillo, S.~Esposito, F.~Iocco, G.~Mangano, G.~Miele and P.~D.~Serpico,
  Comput.\ Phys.\ Commun.\  {\bf 178} (2008) 956


\bibitem{Chluba:2016bvg}
  J.~Chluba,
  arXiv:1603.02496 [astro-ph.CO].
  
\bibitem{Vincent:2014rja} 
  A.~C.~Vincent, E.~F.~Martinez, P.~Hernndez, M.~Lattanzi and O.~Mena,
  JCAP {\bf 1504}, no. 04, 006 (2015)
  doi:10.1088/1475-7516/2015/04/006
  [arXiv:1408.1956 [astro-ph.CO]].

\bibitem{deSalas:2015glj}
  P.~F.~de Salas, M.~Lattanzi, G.~Mangano, G.~Miele, S.~Pastor and O.~Pisanti,
  Phys.\ Rev.\ D {\bf 92},  123534 (2015).

\bibitem{Chikashige:1980ui} 
  Y.~Chikashige, R.~N.~Mohapatra and R.~D.~Peccei,
  Phys.\ Lett.\ B {\bf 98}, 265 (1981).
  
\bibitem{Bazzocchi:2008fh} 
  F.~Bazzocchi, M.~Lattanzi, S.~Riemer-S{\o}rensen and J.~W.~F.~Valle,
  JCAP {\bf 0808}, 013 (2008).

\bibitem[Zdziarski 
\& Svensson(1989)]{1989ApJ...344..551Z} Zdziarski, A.~A., \& Svensson, R.\ 1989, Astrophys.\ J.\ , 344, 551

\bibitem[Kawasaki 
\& Moroi(1995)]{1995ApJ...452..506K} Kawasaki, M., \& Moroi, T.\ 1995, Astrophys.\ J.\ , 452, 506




\end{thebibliography}
\end{document}